# Molecular mechanisms, thermodynamics, and dissociation kinetics of knob-hole interactions in fibrin*

**Olga Kononova[1,2], Rustem I. Litvinov[3], Artem Zhmurov[1,2], Andrey Alekseenko[2], Chia Ho Cheng[1], Silvi Agarwal[1], Kenneth A. Marx[1], John W. Weisel[3], and Valeri Barsegov[1,2]**

[1]Department of Chemistry University of Massachusetts, Lowell, MA 01854

[2]Moscow Institute of Physics and Technology, Moscow Region, Russia 141700

[3]Department of Cell and Developmental Biology
University of Pennsylvania School of Medicine, Philadelphia, PA 19104

*Running Title: *Mechanisms of knob-hole interactions in fibrin*

To whom correspondence should be addressed: Valeri Barsegov, Department of Chemistry, University of Massachusetts, Lowell, MA, USA, Tel.: (978) 934-3661; E-mail: Valeri_Barsegov@uml.edu; John W. Weisel, Department of Cell and Developmental Biology, University of Pennsylvania School of Medicine, Philadelphia, PA, USA, Tel: (215) 898-3573; Fax: (215) 898-9871; E-mail: weisel@mail.med.upenn.edu

**Background:** Knob-hole interactions underlie formation and properties of fibrin polymer, the scaffold of blood clots and thrombi.

**Results:** The structural mechanisms, dissociation kinetics and thermodynamic parameters of the A:a and B:b knob-hole interactions have been determined.

**Conclusion:** The knob-hole bonds are inherently variable and are sensitive to pH and temperature.

**Significance:** Emerging molecular picture offers mechanistic insights into fibrin polymerization.

## ABSTRACT

Polymerization of fibrin, the primary structural protein of blood clots and thrombi, occurs through binding of knobs 'A' and 'B' in the central nodule of fibrin monomer to complementary holes 'a' and 'b' in the γ- and β-nodules, respectively, of another monomer. We characterized the A:a and B:b knob-hole interactions under varying solution conditions using Molecular Dynamics simulations of the structural models of fibrin(ogen) fragment D complexed with synthetic peptides GPRP (knob 'A' mimetic) and GHRP (knob 'B' mimetic). The strength of A:a and B:b knob-hole complexes was roughly equal, decreasing with pulling force; yet, the dissociation kinetics were sensitive to variations in acidity (pH=5-7) and temperature (T=25-37°C). There were similar structural changes in holes 'a' and 'b' during forced dissociation of the knob-hole complexes: elongation of loop I, stretching of interior region, and translocation of the moveable flap. The disruption of the knob-hole interactions was not an "all-or-none" transition, as it occurred through distinct two-step or single-step pathways with or without intermediate states. The knob-hole bonds were stronger, tighter, and more brittle at pH=7 than at pH=5. The B:b knob-hole bonds were weaker, looser, and more compliant than the A:a knob-hole bonds at pH=7, but stronger, tighter, and less compliant at pH=5. Surprisingly, the knob-hole bonds were stronger, not weaker, at elevated temperature (T=37°C) compared to T=25°C due to the helix-to-coil transition in loop I, which helps stabilize the bonds. These results provide detailed qualitative and quantitative characteristics underlying the most significant non-covalent interactions involved in fibrin polymerization.

Formation and decomposition of fibrin clots are essential for hemostasis, thrombosis, and wound healing (1-3). Fibrin network formation is initiated by limited proteolysis of fibrinogen by thrombin, resulting in polymerization of fibrin in two major steps: self-assembly of fibrin monomers into two-stranded half-staggered rod-like protofibrils and lateral aggregation of protofibrils into thicker fibrils that form the branched 3D network (4-7). Building of fibrin protofibrils is driven by the intermolecular A:a knob-hole interactions, while B:b knob-hole bonds are involved in the lateral aggregation of protofibrils. Roughly, the length and diameter of fibrin fibers are determined by the relative rates of longitudinal oligomerization vs. lateral aggregation of fibrin oligomers reaching a critical length (8). During and after formation, the stability of blood clots in response to mechanical forces imposed by the blood flow, wound stretching, and other dynamic environmental conditions is regulated by the unbinding kinetics of the knob-hole interactions until the clot is cross-linked by Factor XIIIa (9). Consequently, the binding and unbinding kinetics of knob-hole interactions determine formation of fibrin fibers, influence the final structure and stability of clots and thrombi, including a potential of clot remodeling, embolization, contraction, and other (patho)physiological processes related to blood clotting and thrombosis. Impaired knob-hole interactions results in loose, weak, unstable clots and are associated with the tendency to bleed. Dense fibrin networks originating from enhanced knob-hole interactions show increased stiffness, a higher fibrinolytic resistance and mechanical resilience, which may predispose individuals to cardiovascular disease such as heart attack and stroke (10-12).

Fibrinogen, the soluble fibrin precursor, consists of three pairs of polypeptide chains, Aα, Bβ, and γ, linked together by 29 disulfide bonds (13). Thrombin splits off two pairs of fibrinopeptides A and B from the N-termini of the Aα and Bβ chains, respectively, in the central nodule. This results in the exposure of binding sites 'A' and 'B' that interact, respectively, with constitutively accessible sites 'a' and 'b' in the γ- and β-nodules of the lateral D regions of another fibrin molecule (Figure 1) (14-16). The polymerization sites have also been called knobs

'A' and 'B' and holes 'a' and 'b' (14) because X-ray crystallographic studies of fibrinogen fragments revealed binding pockets (holes) complementary to the peptides, GPRP and GHRP, corresponding to the newly exposed amino terminal ends (knobs) of the α and β chains of fibrin (17). Since the structure of the actual complexes that form in fibrin polymerization have not been observed, it is not yet known whether the binding sites consist only of the peptides fitting into the holes or the interfaces or the association processes are more complex involving other surface amino acids of the two interacting species.

The N-terminal α chain motif GPR, the main functional sequence in the knob 'A', is complementary to the hole 'a' located in the γ-nodule. The N-terminal β chain motif GHRP is a major part of the knob 'B' that binds to the hole 'b' located in the β-nodule. Analysis of the structures of fragment D (containing the γ-nodule) co-crystallized with GPRP peptide (synthetic knob 'A' mimetic) has revealed that the binding hole 'a' is localized to residues γ337-379 of the γ-nodule: γAsp364, γArg375, γHis340 and γGln329 accommodate binding of the GPRP peptide, and γLys338 and γGlu323 shift slightly to allow γLys338 to interact with the C-terminus of the peptide (Figure 1) (18). Due to homology of the amino acid sequences forming the hole 'a' (in the γ-nodule) and hole 'b' (in the β-nodule) and structural similarity of their binding pockets (Figure 1), the hole 'b' involves similar binding regions βAsp383-Asp398, βTyr404-Gly434, and βGln359-Ile369, which accommodate formation of binding contacts with and subsequent association of the peptide GHRP (synthetic knob 'B' mimetic).

Although the X-ray crystallographic studies have provided valuable structural data about the binding sites mediating the knob-hole associations, this information is limited to a static molecular image of the A:a and B:b knob-hole complexes. Optical trap-based force spectroscopy and atomic force microscopy have been used to probe directly the strength of knob-hole interactions via dissociation of knob-hole bonds at the single-molecule level (6,19,20). Yet, these experiments have limited spatial resolution and do not reveal the molecular mechanisms of the knob-hole interactions. To address these limitations, here we have embarked on the computational exploration of the knob-hole interactions in fibrin using Molecular Dynamics (MD) simulations, which, in conjunction with atomic structural models, help to advance our understanding of protein function and dynamics (21). In computer-based modeling of the force-induced dissociation of protein-protein complexes, conditions of force application are similar to force protocols employed in dynamic force spectroscopy (22,23). To approach physiologically relevant conditions, tensile forces and temperature can be varied, ionic strength can be modeled by including an appropriate number of ions in the solvation box, and the solvent acidity (pH) can be modeled by taking into account the degree of protonation of amino acid residues.

In this work, we have performed Molecular Dynamics (MD) simulations of forced dissociation of non-covalent A:a and B:b bonds to explore and compare the dynamic behavior of the A:a and B:b knob-hole complexes subject to tension. Here, we report the results of our studies of the kinetics (timescales and reaction pathways), thermodynamics (energy landscapes), and molecular mechanisms of the forced dissociation of the A:a and B:b knob-hole complexes performed under different virtual ambient conditions (pH and temperature). We have employed the atomic resolution inherent to MD simulation approaches to assess the importance of particular amino acid residues and clusters of residues for the binding affinity and strength of the knob-hole interactions. We have probed the dynamic network of residues in the holes 'a' and 'b' forming electrostatic contacts with the peptides GPRP and GHRP, respectively. Taken together, the results obtained provide a broad foundation for understanding the key interactions in fibrin polymerization. The kinetic and thermodynamic characteristics can also be used to formulate new drug design strategies to attenuate the knob-hole interactions in fibrin, in order to modify the fibrin clot structure and properties and to reduce the danger of thromboembolic complications.

## EXPERIMENTAL PROCEDURES

**Model systems for A:a and B:b knob-hole complexes** – Structural models for the A:a and B:b knob-hole complexes were obtained using the X-ray structure of double-D fragment from human fibrin containing both holes 'a' and 'b' co-complexed with the Gly-Pro-Arg-Pro-amide (GPRPam) peptide and Gly-His-Arg-Pro-amide (GHRPam) peptides (PDB entry 1FZC) (24). In these structures, holes 'a' and 'b' contain residues $\gamma143$-392 and $\beta197$-458, respectively. We used the synthetic knob 'A' (GPRP) and knob 'B' (GHRP), which mimic the N-terminal portions of knobs 'A' and 'B' binding with holes 'a' and 'b', respectively (18). The hole 'a' is localized to clusters of residues $\gamma Trp315$-Tpr330, $\gamma Trp335$-Asn365 and $\gamma Phe295$-Thr305 in the $\gamma_D$-nodule. The hole 'b' in the $\beta$-nodule involves binding regions $\beta Asp383$-Asp398, $\beta Tyr404$-Gly434, and $\beta Gln359$-Ile369 (Figure 1). Summarized description for preparation of each model system Aa1 (pH=7, T=37°C), Aa2 (pH=7, T=25°C), Aa3 (pH=5, T=37°C), and Aa4 (pH=5, T=25°C) for the A:a knob-hole bond, and Bb1 (pH=7, T=37°C), Bb2 (pH=7, T=25°C), Bb3 (pH=5, T=37°C), and Bb4 (pH=5, T=25°C) for the B:b knob-hole bond is given in Table S1 and S2 in Supplemental Data (SD). To model the pH-dependence of forced dissociation of knob-hole bonds, we considered the degree of protonation of amino acids. Because at pH=5 only His residues are protonated compared to pH=7 ($pK_a \approx 6$), we replaced neutral His residues with positively charged His residues. Ion concentration translates to including an appropriate number of ions $Na^+$ and $Cl^-$ in the water box. We used the relevant physiological 150 mM concentration of NaCl. Details regarding the system preparation are given in SD (Table S1 and S2). Each system was solvated in a water box; the number of water molecules and size of salvation box are given in Table S2 in SD. In Umbrella Sampling calculations, we increased the volume of the water box to 47.6 Å x 49.1 Å x 98.7 Å (for systems Aa1-Aa4) and to 50.1 Å x 50.9 Å x 103.6 Å (for systems Bb1-Bb4). Each model system was first minimized for 5000 steps using the steepest descent algorithm. After initial minimization, each system was heated to T=25°C or 37°C and equilibrated for 0.3 ns. The simulations were carried out at constant temperature. The water density was maintained at 1 $g/cm^3$, and periodic boundary conditions were applied to the water molecules. Non-bonded interactions were switched off at 12 Å distance, and had a switching function from 10 Å to 12 Å. The long-range electrostatics was described using the particle-mesh Ewald method. We used Langevin thermostat to maintain the conditions of constant temperature. The damping coefficient was set to $\gamma = 6\pi\eta a/m = 50$ ps$^{-1}$, which corresponds to the water viscosity $\eta = 0.01$ Poise, and size and mass of amino acid $a = 5\times10^{-8}$ cm and $m = 2\times10^{-22}$ g, respectively.

**MD simulations of the A:a and B:b knob-hole complexes** – To carry out computational modeling of systems Aa1-Aa4 and Bb1-Bb4 we used NAMD 2.7 software package (25) and the CHARMM22 force-field (26). Each initial protein structure was solvated with at least 15 Å of TIP3P water (27) using the VMD solvate plugin (28). We carried out 10-ns equilibrium simulations of the D region of the fibrinogen molecule, which includes the $\gamma$-nodule (with hole 'a') and the $\beta$-nodule (with hole 'b') co-complexed with the peptide GPRP (knob 'A' mimetic) and GHRP (knob 'B' mimetic), respectively (see Figure 1). The forced dissociation of GPRP peptide from the $\gamma$-nodule (A:a knob-hole complex) and GHRP peptide from the $\beta$-nodule (B:b knob-hole complex) was performed using steered Molecular Dynamics implemented in the NAMD package. To mimic the experimental force-clamp measurements, we constrained hole 'a' (hole 'b') by fixing the C-terminal part of one $\gamma$ chain ($\beta$ chain) - residue $\gamma$Gly160 ($\beta$Val205); constant tensile force $\boldsymbol{f} = f\cdot\boldsymbol{n}$ was applied to the $C_\alpha$-carbon of Pro4 residue in peptide GPRP (GHRP) in the direction $\boldsymbol{n}$ perpendicular to the binding interface to dissociate the A:a (B:b) knob-hole bond. We used force $f = 150, 200, 250, 300, 350,$ and $400$ pN. To obtain statistically meaningful information, we performed multiple pulling runs: 15 trajectories of forced unbinding were generated for each force value (total of 90 trajectories for each model system).

**Analysis of simulation output** – *Analysis of Structures:* To probe conformational flexibility of the protein domains forming hole 'a' and hole 'b', we computed the root-mean-square deviations, $RMSD(i) = (1/t_{tot}\sum_{ij}(x_i(t_j) - \tilde{x}_i))^{1/2}$, where $x_i(t_j)$ and $\tilde{x}_i$ are positions of the $C_\alpha$-atom $i$ ($i$-th residue) in a current state and in the reference state, respectively, and $t_{tot}$ is the length of the simulation run. To remove rotational and translational contributions, we used a "running average" structure for each 150-ps time interval as a reference state. To probe the global conformational transitions, we analyzed transient structures for each model system and visualized then using the VMD plugin (28). *Analysis of Kinetics:* We analyzed the average bond lifetimes and standard deviations. A pair of amino acids $i$ and $j$ is said to form a binary contact if the distance between the center-of-mass of their side chains $r_{ij} < 6.5$Å exists for more than 0.1 ns. The time-dependent maps of interacting residues were constructed and the total number of binding contacts $Q(t)$ was used to monitor the dissociation progress. The bond lifetime $\tau$ was defined as the time at which $Q=0$. *Molecular Mechanism:* We utilized Essential Dynamics (29,30) to capture collective displacements of amino acid residues $\wp x(t) = x(t) - x_0$ from their equilibrium positions $x_0$ along the unbinding reaction coordinate $X$ (see SD for detail). We performed numerical diagonalization of the covariance matrix $\boldsymbol{C}(t)= \boldsymbol{M}^{1/2}\wp x(t)(\boldsymbol{M}^{1/2}\wp x(t))^T = \boldsymbol{T}^T \boldsymbol{L} \boldsymbol{T}$ to compute the matrix of eigenvalues $\boldsymbol{L}$ and the matrix of eigenvectors $\boldsymbol{T}$ ($\boldsymbol{M}$ is the matrix of masses of amino acids). These were used to calculate the root-mean-square displacements for each residue $I$ along the eigenmode $k$, i.e. $RMSD^k_I = (C_{II}/M_{II})^{1/2} = (1/M_{II}\sum_{kl}(T^T_{Ik}L_{kl}T_{lI}))^{1/2} = (L_{kl}T_{kl}^2/M_{II})^{1/2}$, in the center-of-mass representation. The native structure of the knob-hole complex was superposed with the structure corresponding to the maximum displacement along the first (principal) mode $k = 1$. *Analysis of Thermodynamics:* To resolve the unbinding free energy landscape $G(X)$, we employed Umbrella Sampling method (31,32), described in detail in the Supplemental Data, to compute the potential of mean force, in order to quantify the interaction energy.

## RESULTS

**Native properties of A:a and B:b knob-hole complexes at various ambient conditions:** To identify the residues in the holes 'a' and 'b' principally involved in dynamic interactions with GPRP and GHRP and assess the plasticity of the knob-hole bonds in response to varying environmental conditions, we explored the A:a and B:b knob-hole complexes in their native state. Since the knob-hole interactions are mainly electrostatic, it should be expected that the strength of knob-hole bonds is susceptible to variation in proton concentration (pH) and temperature, both of which can change at a wound site *in vivo*. It is known that at the sites of inflammation and tumor growth the microenvironment is acidified, e.g., as a result of local metabolic acidosis due to hypoxia (33). Therefore, we have performed equilibrium simulations of the A:a and B:b knob-hole complexes at pH=5 and pH=7, and at T=25°C and T=37°C (body temperature). By varying pH and temperature we generated four model systems to study the A:a knob-hole complexes (Aa1-Aa4) and four model systems for the B:b knob-hole complexes (Bb1- Bb4) described in Experimental procedures and in Supplemental Data (SD) (see Table S1 and S2).

For each of the eight model systems (Aa1-Aa4 and Bb1-Bb4), we analyzed the root-mean-square-deviations (RMSD) quantifying fluctuations for each amino acid residue in holes 'a' and 'b' in the bound state with the corresponding ligand peptide, GPRP and GHRP, respectively. It should be expected that amino acids in holes 'a' and 'b' involved in binding of the peptides would be more flexible to accommodate small-amplitude thermal peptide motion in the binding pocket. Hence, these residues would show larger RMSD values. We found that in all four model systems Aa1-Aa4 for the A:a knob-hole complexes, the largest fluctuations (RMSD>1.5 Å) correspond to the same structural regions in hole 'a' localized to residues γTrp315-Trp330, γTrp335-Asn365, and γPhe295-Thr305 (Figures S1**A** and S1**B**). Specifically, residues γGlu323, γPro360, and γPro299 were found to have the highest RMSD values. These same stretches of residues have been found in earlier studies of the X-ray structures of fragment D co-crystallized with the peptide GPRP to participate in the binding interactions (14,34). These binding determinants were termed loop I or region I (γTrp315-Trp330), interior region or region II (γTrp335-Asn365), and moveable flap or region III (γPhe295-Thr305) (Figures 1**B** and 1**C**). There were smaller changes in RMSD values for amino acid residues in loop I and moveable flap observed at the higher temperature (T=37°C) as compared to the lower temperature (T=25°C) and for the neutral environment (pH=7) as compared to an acidic environment (pH=5). The only significant difference was detected for amino acids in the interior region, for which RMSD values increased from 1.7-1.8 Å to 2.0–2.2 Å upon decreasing pH from 7 to 5 (e.g., see peak at residue γPro360 in Figures S1**A** and S1**B**).

We obtained similar results for the model systems Bb1-Bb4 for the B-b knob-hole complexes (Figures S1**C** and S1**D**). The largest fluctuations detected (RMSD>2.0 Å) correspond to the regions of hole 'b' localized to residues βAsp383-Asp398, βTyr404-Gly434 and βGln359-Ile369. The residues βGly362, βSer388, and βHis429 were found to have the highest RMSD values. Consistent with the previous crystallographic studies (14,34), the three binding regions in hole 'b' had the secondary structure similar to that for loop I (region I), interior region (region II), and moveable flap (region III) forming the binding interface in hole 'a' (compare, e.g., Figures 1**A** and 1**C**, and Figures 1**B** and 1**D**). For this reason, we used these same notations to denote the binding structures in hole 'b': loop I (βAsp383-Asp398), interior region (βTyr404-Gly434), and moveable flap (βGln359-Ile369) (Figure 1**C**). We also found that the mobility of amino acids forming these binding determinants in hole 'b' does not change much upon changing temperature and pH (Figures S1**C** and S1**D**).

To summarize, our results of equilibrium simulations are in full agreement with published X-ray crystallographic data in terms of identifying the binding determinants for the A:a and B:b knob-hole complexes, which validates our MD simulation protocol. The results indicate that the binding structures and near-native ensemble of the A:a and B:b knob-hole bonds are fairly similar. There is a minor difference in mobility of the binding determinants, which are generally more flexible at pH=7 in hole 'b' than in hole 'a'. Varying temperature and pH within the studied ranges does not affect much the binding structures that stabilize the A:a and B:b knob-hole bonds, except for the binding residues γTrp335-Asn365 (region II) in hole 'a', which become more mobile in an acidic environment.

**Forced dissociation kinetics of A:a and B:b knob-hole complexes:** We used a 150-400 pN range of constant pulling force to model the influence of varying stress due to blood flow on the knob-hole bond lifetimes. These forces were chosen as physiologically relevant. The range of hydrodynamic forces of the blood flow acting on knob-hole bonds in fibrin can be obtained using the relationship $f = 6\pi\eta CRhr$, which links the shear rate ($r$) and the tensile force ($f$). Here, $C \approx 1$-2 is the dimensionless constant, $\eta$ is blood viscosity (~10 Poise), $R$ is clot size, and $h$ is the average distance between the clot and vessel walls (~0.1–1 μm). Our estimates show that under (patho)physiological conditions of blood circulation, the knob-hole bonds in fibrin polymers forming $R \approx 10^1$–$10^2$ μm clots are subjected to tensile forces from a few tens of pico-Newtons (normal arterial blood flow with $r \approx 10^2$ s$^{-1}$) to a few nano-Newtons (stenotic blood flow with $r \approx 10^4$ s$^{-1}$).

We mechanically tested the strength of the A:a and B:b knob-hole bonds under varying conditions, i.e. for different temperature (25°C and 37°C) and acidity (pH=5 and pH=7), using pulling simulations (see Experimental procedures). Lowering pH from pH=7 to pH=5 results in the protonation of six His residues ($pK_a \approx 6$) in hole 'a', four His residues in hole 'b', and one His residue in the GHRP peptide (see Table S1 in SD). These additional positive charges could alter the pattern of electrostatic interactions.

The strength of non-covalent bonds usually decreases with increased pulling force, meaning that protein-ligand complexes dissociate faster at higher forces (34-36). We found that as the pulling force ($f$) increased, the average bond lifetimes ($\tau$) decreased and the dependence of $\tau$ on $f$ was monotonic for all model systems, Aa1–Aa4 and Bb1–Bb4 (Figure 2). For equivalent ambient conditions, the A:a and B:b knob-hole bonds were found to be roughly equally strong in the entire force range 150-400 pN (Figure 2). Yet, both A:a and B:b knob-hole bonds were stronger and had longer lifetimes in neutral solution (pH=7) as compared to acidic solution (pH=5). Also, the A:a knob-hole bonds were found to be stronger at T=37°C as compared to T=25°C both for pH=5 and pH=7, whereas the B-b knob-hole bonds were weaker at T= 37°C for pH=7 and almost equally strong at T=37°C and 25°C for pH=5 (Figure 2). Hence, increasing proton concentration has a more profound effect on the A:a and B:b knob-hole bonds as compared to varying temperature.

**Single- and two-step dissociation of A:a and B:b knob-hole complexes:** Statistics of bond lifetimes indicate that the force-driven dissociation of the A:a and B:b knob-hole bonds is intrinsically stochastic and quite variable. This can be seen, e.g., from large standard deviations of the average bond lifetimes, especially at lower 150–250 pN force (Figure 2). To understand the origin of large fluctuations in the bond lifetimes, we analyzed the dynamics of forced disruption of the binding contacts. In the studies of the near-native ensemble, we have identified 16 most important amino acid residues in the binding interface in holes 'a' and 'b', which formed ~90% of all stable contacts with the residues in the GPRP and GHRP peptides, respectively. These are residues γPhe295, γAsp297, γAsp298, γAsp301, γPhe304, γThr305, γPhe322, γCys326, γGln329, γAsp330, γLys338, γCys339, γHis340, γTyr363, γAsp364, γArg375 in hole 'a', and residues βGln359, βLeu360, βAsn364, βThr368, βTrp385, βLys392, βCys394, βGlu397, βAsp398, βArg406, βCys407, βHis408, βThr431, βAsp432, βMet438, βSer443 in hole 'b'.

Next, we performed extensive analyses of the simulation output generated at the lowest attainable pulling force $f$=150 pN. To study the dissociation dynamics, we monitored the total number of stable contacts at the knob-hole interface as a function of time $Q(t)$ (see Experimental procedures), which reflects the instantaneous changes in the strength of the bond during forced unbinding. The time-dependent profiles of $Q$, obtained from the most representative knob-hole unbinding trajectories for the model systems Aa1−Aa4 and Bb1−Bb4, show that $Q(t)$ starts off from ~15-18 contacts for the A:a knob-hole complexes (Figures 3**A** and 3**B**) and from ~15-17 contacts for the B:b knob-hole complexes (Figures 3**C** and 3**D**), which correspond to the native bound state ($B$), but decays to zero along the unbinding pathway. The moment of time $\tau$, at which all the binding contacts have been disrupted and, hence, $Q(t)$=0, signifies a complete dissociation of the complex and formation of the unbound (dissociated) state ($U$). A closer analysis revealed that in some simulation runs, the forced dissociation occurred in a single step, $B \rightarrow U$, whereas in other runs, the dissociation occurred by populating the intermediate state ($I$), i.e. $B \rightarrow I \rightarrow U$. This is reflected in the profiles of $Q(t)$ (Figure 3), some of which show a one-step transition, i.e. initial plateau followed by a sharp decay to zero, whereas other profiles exhibit two-step transitions. The former pathway (P1) corresponds to the single-step transition ($B \rightarrow U$), whereas the latter pathway (P2) represent the two-step unbinding transition ($B \rightarrow I \rightarrow U$). We have estimated the extent of this kinetic partitioning. In the case of the A:a and B:b knob-hole bonds at pH=7, the two-step unbinding pathway P2 was observed in only ~10% of simulated trajectories (Figures 3**A** and 3**C**); in the acidic environment (pH=5) the share of this minor pathway increased to ~30% (Figures 3**B** and 3**D**).

In summary, our results indicate that dissociation of the A:a and B:b knob-hole bonds occurs through multiple competing kinetic pathways with or without intermediate (partially dissociated) states, which can underlie the remarkable variability of bond lifetimes.

**Dynamic maps of binding contacts forming A:a and B:b knob-hole bonds:** To get insight into structural details of the knob-hole formation and dissociation, we constructed and analyzed the time-dependent molecular maps of amino acid residues in holes 'a' and 'b' coupled to the residues in GPRP and GHRP peptides, respectively, during their forced unbinding. Figures S2 and S3 show the maps of stable binary contacts and how they change over time for model systems Aa1–Aa4 and Bb1–Bb4. For all systems Aa1–Aa4 and Bb1–Bb4, stable binary contacts formed between residues in the hole 'a' and the knob 'A' and between residues in the hole 'b' and the knob 'B' are summarized in Table I.

The results obtained indicate that for the A:a knob-hole bonds at pH=7, the strongest contacts are formed between amino acid residues in loop I and residues Gly1 and Arg3 in GPRP; between residues in interior region and residues Gly1, Pro2, Arg3 in GPRP; and between residues in moveable flap and residues Pro2 and Pro4 in GPRP (Figure 1). The pH lowering results in weakening of interactions between residues in interior region and residues Gly1 and Pro2 in GPRP and between residues in loop I and residue Gly1 in GPRP, but also results in formation of stable contacts between Gly1 and Pro2 in GPRP and residues in moveable flap. At pH=5, there is a preference for association with the moveable flap, which is opposite to what is observed at pH=7 where binding to interior region and loop I is more favorable. Redistribution of binding contacts leads to weakening of the A:a knob-hole bonds, also reflected by the decrease of the average bond lifetime for systems Aa3 and Aa4 (Figure 3**A**). This finding implies that molecular interactions of hole 'a' with residues Gly1 and Pro2 in GPRP play a crucial role. For the B:b knob-hole bonds, the strongest contacts are between residues in loop I and interior region and residues Gly1 and Arg3 of GHRP (Figure 1). At pH=7, additional contacts form between His2 of GHRP and residues in interior region and moveable flap. The pH lowering results in weakening of the interactions between residues in interior region and the moveable flap with His2 of GHRP, but preserving contacts between residues in loop I and the interior region with residues Gly1 and Arg3 in GHRP.

**Structural transitions underlying forced dissociation of A:a and B:b knob-hole complexes:** We employed the Essential Dynamics approach (see Experimental procedures; see also SD) to single out the most important types of motion showing the largest contribution in the direction of the global transition (dissociation). Using positions of amino acid residues as a function of time, $\boldsymbol{x}(t) = \{x_1(0), x_2(0),..., x_N(0)\}$, we numerically diagonalized the covariance matrix of their mass-weighted displacements $\Re\boldsymbol{x}(t)$ from the reference (native) structure $\boldsymbol{x}_0=\{x_1(0), x_2(0),..., x_N(0)\}$ (bound state) to obtained the matrix of eigenvalues $\boldsymbol{L}$, and the matrix of eigenvectors $\boldsymbol{T}$. The eigenvalues $L_I$ provide information about the amplitude of the $I$-th eigenvector $t_I$ along the displacement $\Re\boldsymbol{x}(t)$. We calculated the root-mean-square displacement (RMSD) for each $I$-th residue along mode $k$, $RMSD_I^k = (L_{kI}T_{kI}^2/M_{II})^{1/2}$, where $M_{II}$ is the element of $\boldsymbol{M}$ − the matrix of masses of amino acids. The RMSD profiles for the first three modes ($k$=1, 2, and 3) capturing ~85% of the dissociation dynamics for model systems Aa1–Aa4 and Bb1–Bb4 were determined (Figures S4 and S5).

The peaks of $RMSD_I^k$ correspond to residues forming the main binding determinants − loop I, interior region, and moveable flap described in Figure 1 (see also Figure S1). For these regions, the $RMSD_I^k$ values are in the 2–4 Å range for the A:a knob-hole bonds (Figure S4) and in the 2–5 Å range for the B:b knob-hole bonds (Figure S5). The $RMSD_I^k$ values for the A:a and B:b knob-hole bonds are close, implying similar flexibility of their binding interface. At pH=7, the interfaces in the A:a and B:b knob-hole complexes show more flexibility at higher temperature T=37°C (Figures S4**A**, S4**B**, and S5), but when the pH is lowered to 5, the amplitude of motions is roughly the same at T=25°C and 37°C (Figures S4**C**, S4**D**, and S5). For the A:a knob-hole bond, $RMSD^k_I$=3-5 Å for interior region and moveable flap at pH=7 (both at T=25°C and 37°C) and $RMSD^k_I$=2 Å for loop I. For the A:a knob-hole bonds, pH lowering resulted in smaller values of $RMSD^k_I$ for all three regions: at T=37°C motions of binding residues in loop I are quenched (Figure S4). For the B:b knob-hole bonds, changing pH from 7 to 5 leads to increased $RMSD^k_I$=5 Å for residues in moveable flap (Figure S5).

For each system, the most essential structural changes associated with the first (principal) mode (k=1) are summarized in Figure S6. Both for the A:a and B:b knob-hole complexes, dissociation is associated with partial opening of the binding interface, which is accompanied by the simultaneous stretching of all binding regions, the extent of which is determined by ambient conditions. The binding interface in the B:b knob-hole complex is slightly more flexible. The binding determinants in hole 'b' maintain their secondary structure (Figure S6, **E-H**); yet there are minor changes in the secondary structure propensities for regions I and III in hole 'a' (Figure S6, **A-D**). Both for the A:a and B:b knob-hole complexes, there is a displacement of loop I and translocation of moveable flap from the periphery to the center of the interface. The latter transition is coupled to partial stretching of the interior region. Both temperature and pH variation bring about rather small changes to the molecular arrangement.

**Thermodynamics of forced dissociation of A:a and B:b knob-hole bonds:** We resolved the profile of the Gibbs free energy change for forced unbinding, $\mathcal{G}$, as a function of the "receptor-ligand distance", $X$, for various solution conditions (see Experimental procedures; see also SD) (31,32). For the A:a knob-hole complex, $X$ was taken to be the distance between the residue γGly160 in hole 'a' and the C-terminal Pro4 residue in GPRP. For the B:b knob-hole complex, $X$ was taken to be the distance between the residue βVal205 in hole 'b' and the C-terminal Pro4 residue in GHRP. Residues γGly160 and βVal205 were chosen because they are in the core of the binding pockets in holes 'a' and 'b', respectively. The results are presented in Figure 4, where we have compared the profiles of the average $\mathcal{G}$ as a function of the distance change $\Delta X$ for systems Aa1–Aa4 and Bb1–Bb4. We also estimated the "bond width" $\mathcal{x}$ and the "transition state position" $\mathcal{x}^{\oslash}$ directly from the curves of $\mathcal{G}$. These characteristics ($\mathcal{x}$ and $\mathcal{x}^{\oslash}$) are difficult, if not impossible, to obtain experimentally. The average width of the bond $\mathcal{x}$, which we have defined as half of the width of the energy well for the bound state (B) corresponding to the energy change larger than thermal fluctuations ($\mathcal{G}$>$3k_BT$≈1.8 kcal/mol at T=25°C and 1.9 kcal/mol at T=37°C), shows how tightly coupled the protein and ligand molecules are. For example, a steeper growth in $\mathcal{G}$ and, hence narrower $\mathcal{x}$, is indicative of a stronger and tighter coupling. The position of the transition state $\mathcal{x}^{\oslash}$, or the critical bond extension at which $\mathcal{G}$ reaches the energy plateau (bond dissociation), is a measure of the conformational tolerance of the bimolecular complex under tension. Smaller (larger) values of $\mathcal{x}^{\oslash}$ correspond to more brittle (more flexible) bonds. The values of these parameters ($\mathcal{x}$, and $\mathcal{x}^{\oslash}$) and the bond dissociation energy $G_b$, defined as the plateau of $\mathcal{G}$ at large $\Delta X$ (see Figure 4**A**), for all model systems Aa1–Aa4 and Bb1–Bb4 are summarized in Table II.

For all the systems studied, the curves of $\mathcal{G}$ show a well around the minimum at a zero bond extension ($\Delta X$=0), but $\mathcal{G}$ rapidly increases, reaching an energy plateau at larger values of $\Delta X$=1–2 nm. The smooth curves of $\mathcal{G}$ for the A:a and B:b knob-hole complexes, obtained in the neutral environment (pH=7), correspond to the single-step transition (B → U), which is the dominant pathway of dissociation P1 described earlier (Figure 4). Yet, the curves of $\mathcal{G}$ obtained in the acidic environment (pH=5) show several shallow energy wells separated by regions of energy increase. This is because $\mathcal{G}$ represents an ensemble average picture and, hence, it also contains the contribution from the two-step transition (B → I → U), the alternative minor pathway of dissociation P2 (Figure 4). Both the A:a and B:b knob-hole bonds are characterized by more narrow $\mathcal{x}$=0.15-0.19 nm for A:a knob-hole bond and $\mathcal{x}$=0.22-0.29 nm for the B:b knob-hole bond and shorter $\mathcal{x}^{\oslash}$ =0.98-0.99 nm for A:a knob-hole bond and $\mathcal{x}^{\oslash}$ =1.21-1.31 nm for the B:b knob-hole bond at pH=7, as compared to the same quantities at pH=5, i.e. $\delta x$ =0.41-0.55 nm $\delta x^{\oplus}$ =1.45-1.51 nm for A:a knob-hole bond and $\delta x$ =0.30-0.45 nm, $\delta x^{\oplus}$ =1.25-1.37 nm for the B:b knob-hole bond (Table II). Also, the A:a knob-hole bonds are characterized by smaller $\delta x$ and shorter $\delta x^{\oplus}$ than the B:b knob-hole bonds at pH=7, but larger $\delta x$ and longer $\delta x^{\oplus}$ at pH=5 (Table II). The temperature variation does not seem to affect much the values of $\delta x$ and $\delta x^{\oplus}$.

To summarize, our results reveal that both the A:a and B:b knob-hole bonds are stronger in the neutral solution (pH=7) compared to the acidic environment (pH=5). The A:a knob-hole bonds are stronger than the B:b knob-hole bonds at pH=7 but weaker at pH=5 (Figure 4). These conclusions become evident when comparing for all model systems the values of bond dissociation energy $G_b$ (Table II). Quite unexpectedly, the A:a and B:b knob-hole bonds become stronger, not weaker, upon the temperature increase from 25°C to 37°C both in acidic and neutral solution. The A:a and B:b knob-hole bonds tend to form tighter and more brittle complexes at pH=7 than at pH=5. The B:b knob-hole bonds are looser and more compliant than the A:a knob-hole bonds at pH=7, but tighter and more brittle at pH=5.

## DISCUSSION AND CONCLUSIONS

Fibrin polymerization driven by knob-hole interactions is a highly dynamic, kinetically controlled process (37). From the previous X-ray crystallographic studies, knowledge about the molecular interactions mediating formation of the A:a and B:b knob-hole complexes has been limited to the structure of fragment D co-crystallized with synthetic peptides, GPRP and GHRP, mimicking knobs 'A' and 'B', respectively (17,18,38). On the other hand, the state-of-the-art experimental instrumentation, such as optical trap-based force spectroscopy, have made it possible to quantify directly the strength of the A:a and B:b knob-hole bonds at the single-molecule level (6,19). Yet, these experiments cannot access the full dynamics of molecular transitions and resolve the molecular structural details of coupling of the knobs 'A' and 'B' to the holes 'a' and 'b', respectively. The heterogeneity of experimental force signals, partial elongation of fibrin molecules, and the presence of non-specific interactions all make interpretation of experimental data difficult. Despite the critical biological and clinical significance of blood clotting, there have been no theoretical studies of the kinetics (timescales, pathways) and thermodynamics (energy landscape) of the knob-hole interactions aiming at a fundamental understanding of their mechanism(s) at the molecular and sub-molecular level. Here, we showed that these goals can be achieved by using Molecular Dynamics (MD) simulations, which, in combination with experimental results, continue to play an important role in advancing our understanding of the protein-protein interactions (39). In recent years, dynamic force measurements *in silico*, in which tensile forces are used to unfold proteins and to dissociate protein-protein complexes, have become a powerful tool to interpret and clarify the results of nanomechanical experiments *in vitro* (22,39,40).

Here, for the first time, using a combination of MD simulations and theoretical methods, we explored the kinetics and thermodynamics and resolved the structural mechanisms of the knob-hole interactions in fibrin, which initiate and drive fibrin polymerization. These approaches have enabled us to describe in atomic detail the native properties of the A:a and B:b knob-hole complexes and their force-induced dissociation. We also probed the influence of varying important ambient conditions, pH and temperature, on the A:a and B:b knob-hole coupling. The rationale for varying these parameters is that the knob-hole interactions are mainly electrostatic and, hence, are susceptible to a change in temperature and acidity, both of which have (patho)physiological significance. Our knowledge regarding the mechanism of knob-hole interactions in fibrin would be incomplete had we not understood the dynamic mosaic of binding residues stabilizing the A:a and B:b knob-hole bonds. For this reason, we constructed and analyzed the entire molecular maps of amino acid residues in holes 'a' and 'b', which establish strong persistent binding contacts with residues in the peptides GPRP (knob 'A' mimetic) and GHRP (knob 'B' mimetic), respectively. This has helped us identify the residues critical for binding and assess the relative importance of specific amino acid residues, clusters of residues, and even whole binding determinants. These comprehensive efforts have enabled us to extract qualitative and quantitative characteristics of fibrin polymerization.

Profiling the dependence of the knob-hole bond lifetimes on a tensile force revealed that the A:a and B:b knob-hole bonds are roughly equally strong when probed mechanically, albeit the dissociation kinetics are sensitive to pH and temperature variation (Figure 3). The finding that the strength of the A:a and B:b knob-hole bonds is similar is at odds with the recently reported experimental data, according to which the A:a knob-hole bond is about six-fold stronger than the B:b knob-hole bond (6,19). To shed light on the origin of this disagreement, we performed pilot MD simulations by reproducing formation of the A:a knob-hole bonds using not just a GPRP-hole 'a' construct, but a short double-stranded fibrin oligomer formed by three fibrin monomers. In short, the oligomer was first built by superposing the structures of the double-D fragment (PDB code: 1FZC (24)) and fibrin monomer (PDB code: 3GHG (41)); the unresolved portions of the α-chain and β-chain with knobs 'A' and 'B' were reconstructed manually (Figure 5**A**). Next, we ran long 250 ns simulations for the oligomer. We found that binding of the N-terminal end of the α chain (knob 'A') to the pocket in the γ-nodule (hole 'a') is accompanied by electrostatic coupling between residues γGlu323, γLys356, and γAsp297 in the γ-nodule and residues βLys58, βAsp61, and βHis67 in the central nodule of the adjacent fibrin monomers (Figures 5**B** and 5**C**). Hence, our preliminary data seem to indicate that the binding interface might extend beyond the GPR motif exposed upon thrombin cleavage at the N-termini of the α chains, which is traditionally named knob 'A', but more simulations are needed to verify this result. In fact, secondary involvement of the N-terminal portion of the β chain in the A:a knob-hole binding is consistent with transient interactions between fibrin molecules mediated by the N-terminus of the β chain revealed at the single-molecule level (42). Furthermore, this finding supports an idea that the A:a knob-hole interactions in fibrin might have a much broader interface than just the peptide-in-pocket complex. The difference in the binding interfaces between natural protein complexes and peptide-based constructs is the most likely source of disagreement between the published experimental data and our results regarding the relative strength of the A:a and B:b knob-hole bonds. To the best of our knowledge, this is the first substantial evidence for the interactions mediated by "the full knob", which extends beyond the N-terminal GPR sequence.

Although the A:a (and, perhaps, B:b) knob-hole interactions are very likely not limited to the GPR (or GHRP) motifs, the corresponding peptides had been widely used as "synthetic knobs" and as effective competitive inhibitors of the knob-hole interactions (6,9,19, 42-45). Thus, the core of the knob-hole binding characterized using X-ray crystallography is represented by the complexes formed with the GPRP and GHRP peptides, which has been successfully used to study the most basic aspects of the A:a and B:b knob-hole interactions (45,46).

Another important finding is that the A:a and B:b knob-hole interactions in fibrin are not the end product of "all-or-none" transitions as they might occur through distinct pathways via formation of intermediate states (pathway P2 in Figure 3). An immediate consequence of this finding in the context of fibrin polymerization is that this might lead to the formation of the A:a and B:b knob-hole bonds of varying strength. This should be expected since the knob-hole interactions occur in the environment where the contact duration and tensile force change due to the varying blood shear. There is a question whether the knob-hole bonds in fibrin exhibit the so-called "catch-slip" behavior (34,35,47,48), i.e. when the strength of the bond, quantified by the average lifetime, first increases with increasing tensile force and then decreases at higher forces. This unusual type of protein-ligand interaction has been demonstrated for a number of interacting pairs, such as such as P-selectin/PSGL-1 (34), GP1bα/von Willebrand factor (49), bacterial adhesin FimH/mannose (50), integrin α5β1/fibronectin (51), and integrin LFA-1/ICAM-1 (52). Our results show that the A:a and the B:b knob-hole bonds behave as typical "slip" bonds in the 150–400 pN range of tensile forces, but we do not rule out the possibility that the knob-hole bonds might behave as "catch" bonds at lower forces (<150 pN). Because pulling simulations become prohibitively more expensive computationally at lower pulling forces, we were unable to probe the force-dependence of the strength of the A:a and B:b knob-hole bonds below 150 pN.

Our results indicate that despite some differences in the kinetics depending on the magnitude of applied force and variation in temperature and pH, there are similar structural transitions in holes 'a' and 'b', which accompany the force-induced dissociation of the A:a and B:b knob-hole bonds. We singled out the most important modes of motion in the direction of the "reaction coordinate" - receptor-ligand distance, both type and amplitude, which contribute the most to the forced dissociation reaction. We found that for the A:a (B:b) knob-hole bonds, these structural changes are the following: 1-4 Å (2-4 Å) elongation of loop I, stretching of the interior region by 3.5–4.5 Å (1–4 Å), and 2.5–7 Å (2–6.5 Å) translocation of the moveable flap. The extent of these changes depends on the degree of protonation and temperature (Figure S6), and the amplitude of motions is larger at higher temperature.

Because in the simulations we have maintained the conditions of constant pressure and temperature, we used the Gibbs free energy to describe the thermodynamics of the A:a and B:b knob-hole interactions in fibrin. The profiles of the Gibbs free energy change, $\Delta G$, as a function of the change in the interaction distance, $\Delta X$, indicate rather strongly that both in the neutral solution and acidic solution the A:a and B:b knob-hole bonds become stronger at higher temperature T=37°C compared to the lower temperature T=25°C. For the A:a knob-hole bonds, for pH=7 $G_b$=19.3 kcal/mol at 37°C and 16.2 kcal/mol at 25°C and for pH=5 $G_b$=6.1 kcal/mol at 37°C and 1.7 kcal/mol at 25°C. For the B:b knob-hole bonds, for pH=7 $G_b$=15.3 kcal/mol at 37°C and 12.6 kcal/mol at 25°C and for pH=5 $G_b$=9.2 kcal/mol and 8.4 kcal/mol at 25°C (see Table II). This is counterintuitive given that, in general, thermal fluctuations tend to destabilize and weaken non-covalent bonds. To find a structural basis for these unusual findings, we have compared the output from Umbrella Sampling calculations at T=37°C and T=25°C and have found that, both in hole 'a' and hole 'b' in neutral and acidic solution, there is the $\alpha$-helix-to-random-coil transition in loop I, which occurs at a higher temperature (T=37°C). The "melting transition" is displayed for the A:a knob-hole bond in Figure 6, where we have compared the structures of the A:a knob-hole complex at 37°C and 25°C (pH=7; model systems Aa1 and Aa2). We see that residues $\gamma$327-330 in loop I form an $\alpha$-helical pitch at 25°C, which melts into a more flexible random coil structure at 37°C. As a result, residues $\gamma$Glu328, $\gamma$Gln329, and $\gamma$Asp330 come closer and bind stronger with Arg3 in GPRP, which provides an additional ~2.5–4.5 kcal/mol stabilization to the A:a knob-hole bond. This combined entropic effect (order-disorder transition) and enthalpic effect (formation of stronger contacts) compensates for the thermal destabilization of the knob-hole interfaces and accounts for the increased stability of the A:a knob-hole bond at T=37°C as compared to T=25°C. We found a similar transition in the loop I in hole 'b' (data not shown).

To conclude, we have performed a comprehensive study of the molecular mechanisms, thermodynamics, and kinetics of the knob-hole interactions in fibrin using theory and simulations. The results obtained provide a broad theoretical foundation for the key interactions in the fibrin polymerization process, and offer physiologically relevant structural mechanistic insights into this biologically important process at the molecular and sub-molecular level. The results of these studies can be potentially helpful in translational research aiming at the computer-based design of fibrin-specific compounds (53) that could attenuate the knob-hole interactions in a desired fashion, or modify the final clot structure so as to reduce the risk of thromboembolic complications.

*Acknowledgements* – This work was supported by the American Heart Association (Grant 09SDG2460023 to V.B.), Russian Ministry of Education and Science (Grant 14.A18.21.1239 to V.B.), and by the National Institutes of Health (Grants HL030954 and HL090774 to J.W.W.)

**FIGURE LEGENDS**

**FIGURE 1.** Ribbon structures of fibrin(ogen) (panel **A**), the A:a knob-hole bond (panels **B** and **C**), and the B:b knob-hole bond (panels **D** and **E**). The structures correspond to the A:a knob-hole complex (model system Aa1) and B:b knob-hole complex (system Bb1), respectively, at pH=7 and T=25°C. Panels **B** and **D**: the interface of the A:a knob-hole complex (panel **B**) and B:b knob-hole complex (panel **D**), in which the binding determinants – loop I (region I shown in blue), interior region (region II shown in green), and moveable flap (region III shown in red) interact with peptides GPRP and GHRP (shown in orange), respectively. Panels **C** and **E**: Simulation setup: the holes 'a' and 'b' are constrained through fixing the C-termini of the γ chain (residue γGly160) and β chain (residue βVal205), respectively (see Experimental procedures ). A constant pulling force $f$ (represented by the black arrow) is applied to the Pro4 residue of GPRP peptide and Pro4 residue of GHRP peptides in the direction perpendicular to the binding interface to dissociate the knob-hole bond. Also shown are structural details of A:a and B:b knob-hole bonds, in which residues in binding regions I-III in holes 'a' and 'b' establish binding contacts with peptides GPRP and GHRP.

**FIGURE 2.** Kinetics of the forced dissociation of the A:a and B:b knob-hole complexes. The average bond lifetimes ($<\tau>$) with standard deviations for the A:a knob-hole complex (model systems Aa1-Aa4; panel **A**) and for the B:b knob-hole complex (model systems Bb1-Bb4; panel **B**) as a function of pulling force ($f$) are compared for different ambient conditions (pH=5 and pH=7 and T=25°C and T=37°C; see Tables S1 and S2 in SD).

**FIGURE 3.** Dependence of kinetic pathways for forced dissociation of the A:a and B:b knob-hole bonds on pH and temperature. Shown are the time-dependent profiles of the total number of binary contacts ($Q$) stabilizing the A:a knob-hole complex for model systems Aa1 and Aa2 (panels **A**), and Aa3 and Aa4 (panel **B**), and stabilizing the B:b knob-hole complex for model systems Bb1 and Bb2 (panels **C**), and Bb3 and Bb4 (panel **D**). The profiles of $Q$ indicate two distinct dissociation pathways, the one-step pathway of unbinding ($B \rightarrow U$) from the bound state ($B$) to the unbound state ($U$) and the two-step pathway of unbinding ($B \rightarrow I \rightarrow U$), in which formation of the intermediate state ($I$) occurs. The time-dependent maps of binary contacts for A:a and B:b knob-hole bond complexes for different pH values and temperature are presented in Figure S2 and S3 in SD, respectively.

**FIGURE 4.** Free energy landscape underlying the thermodynamics of A:a and B:b knob-hole interactions in fibrin. The Gibbs free energy for unbinding $\Delta G$ for model systems Aa1 and Aa2 (panel **A**), and Aa3 and Aa4 (panel **B**), and for model systems Bb1 and Bb2 (panel **C**), and Bb3 and Bb4 (panel **D**) as a function of knob-hole interaction range $X$ are compared for different ambient conditions (pH=7 and pH=5, and T=25°C and 37°C; see Tables S1 and S2 in SD). The standard deviations of $\Delta G$ are shown for the knob-hole distance change $\Delta X$ = 0.25, 0.5, 0.75, 1.0, 1.25, 1.5, and 1.75 nm for the model systems Aa1 and Aa3 (for the A:a knob-hole bond), and Bb1 and Bb3 (for the A:a knob-hole bond). The standard deviations for systems Aa2 and Aa4, and Bb2 and Bb4 are even smaller. The values of the equilibrium binding energy $G_b$, the width of the bound state $\Delta x$, and the distance between the bound state and transition state $^\ddagger x^\ominus$, shown in panel **A**, are given in Table II.

**FIGURE 5.** Computational reconstruction of the non-covalent coupling of the central nodule (bearing sites 'A') and the γ-nodules (bearing sites 'a'). Panel **A**: ribbon representation of the initial structure (before equilibration) of the double-D fragment of abutted fibrin molecules containing two γ- and two β-nodules. The residues in site 'a' form binding contacts with the residues of site 'A' emanating from the central nodule of the third fibrin monomer between the coiled-coil connectors. Panel **B** shows the translocation of the central nodule following formation of the A:a knob-hole bonds observed at the end of the simulation run. Also shown is the magnified view of electrostatic contacts between residues γGlu323 in loop I, γLys356 in interior region, and γAsp297 in moveable flap (all residues belong to site 'a' in the γ-nodule), and residues βLys58, βAsp61, and βHis67 in the N-terminal portion of the β chain in the central nodule (GPR motif has been suppressed for clarity).

**FIGURE 6.** Comparison of the A:a knob-hole interactions in neutral solution (pH=7) at T=37°C (model system Aa1; panel **A**) and T=25°C (model system Aa2; panel **B**). Shown are the ribbon structures of the binding site 'a' interacting with the knob 'A'. Color denotation: in hole 'a' α-helices are shown in red color, β-strands are in blue color, and coils and turns are shown in grey color; knob 'A' is displayed in green color. Residues γ327-330 in loop I form an α-helix at 25°C, but transition to a random coil structure at 37°C. Interacting residues in loop I and GPRP are magnified below. The electrostatic coupling between residue γTyr363, residues γGlu328, γGln329, and γAsp330 in loop I, and residue Arg3 in the GPRP peptide is indicated.

**TABLE I.** Stable binary contacts between amino acid residues in the hole 'a' and the knob 'A', and between residues in the hole 'b' and the knob 'B', which stabilize the non-covalent A:a and B:b knob-hole bonds in fibrin.

| Model systems | Residues involved in A:a and B:b knob-hole interactions | |
|---|---|---|
| | Holes 'a' and 'b' | Knobs 'A' and 'B' |
| Aa1 and Aa2 (pH=7) | γPhe295, γAsp330, γCys339, γHis340, γArg375 | Gly1 in GPRP |
| | γPhe295, γAsp297, γAsp298, γThr305, γAsp364, γArg375 | Pro2 in GPRP |
| | γPhe322, γCys326, γGln329, γAsp330, γCys339, γTyr363 | Arg3 in GPRP |
| Aa3 and Aa4 (pH=5) | γPhe295, γAsp297, γThr305, γAsp364, γArg375 | Gly1 in GPRP |
| | γPhe295, γAsp297, γAsp298, γAsp301, γThr305 | Pro2 in GPRP |
| | γPhe322, γCys326, γGln329, γAsp330, γCys339, γTyr363, γAsp364 | Arg3 in GPRP |
| Bb1 and Bb2 (pH=7) | βAsp398, βCys407, βHis408, βAsp432, βMet438, βSer443 | Gly1 in GHRP |
| | βLeu360, βAsn364, βThr308, βMet438, βSer443 | His2 in GHRP |
| | βCys394, βGlu397, βAsp398, βCys407, βThr431 | Arg3 in GHRP |
| Bb3 and Bb4 (pH=5) | βAsp398, βCys407, βHis408, βAsp432, βMet438, βSer443 | Gly1 in GHRP |
| | βLeu360, βAsn364 | His2 in GHRP |
| | βCys394, βGlu397, βAsp398, βCys407 | Arg3 in GHRP |

**TABLE II.** Average thermodynamic parameters (binding energy $G_b$, width of the bound state $\wp x$, and transition state position $\wp x^\oplus$ with the standard deviations for the A:a and B:b knob-hole interactions obtained for different model systems using the Umbrella Sampling calculations (see Experimental procedures; see also SD and Tables S1 and S2).

| Model system | $G_b$, kcal/mol | $\wp x$, nm | $\wp x^\oplus$, nm |
|---|---|---|---|
| Aa1 | 19.3 | 0.15 | 0.98 |
| Aa2 | 16.2 | 0.19 | 0.99 |
| Aa3 | 6.1 | 0.55 | 1.45 |
| Aa4 | 1.7 | 0.41 | 1.51 |
| Bb1 | 15.3 | 0.29 | 1.31 |
| Bb2 | 12.6 | 0.22 | 1.21 |
| Bb3 | 9.2 | 0.45 | 1.25 |
| Bb4 | 8.4 | 0.30 | 1.37 |

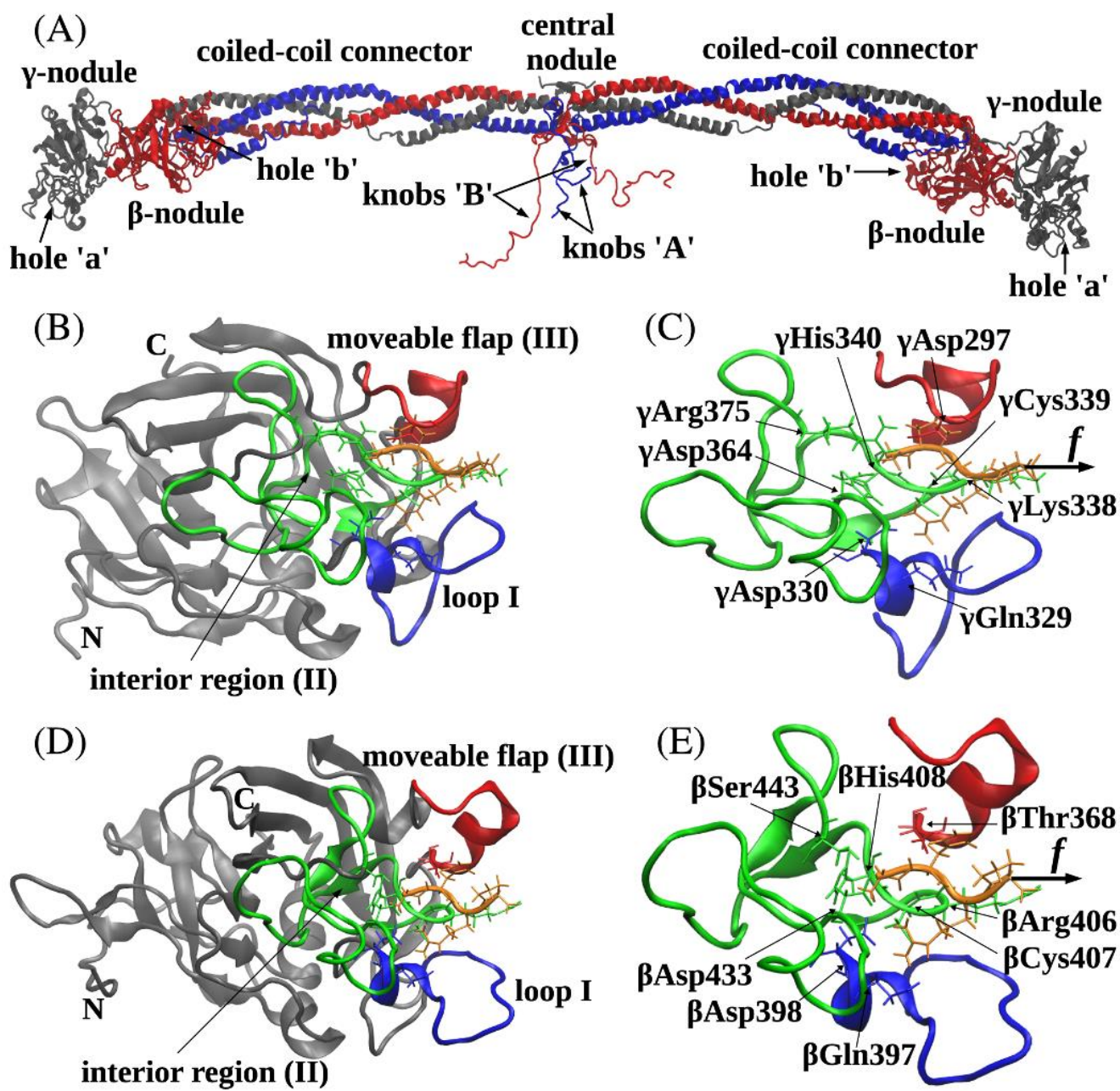

**Figure 1 (Kononova, Litvinov, Zhmurov, Alekseenko, Cheng, Agarwal, Marx, Weisel, Barsegov)**

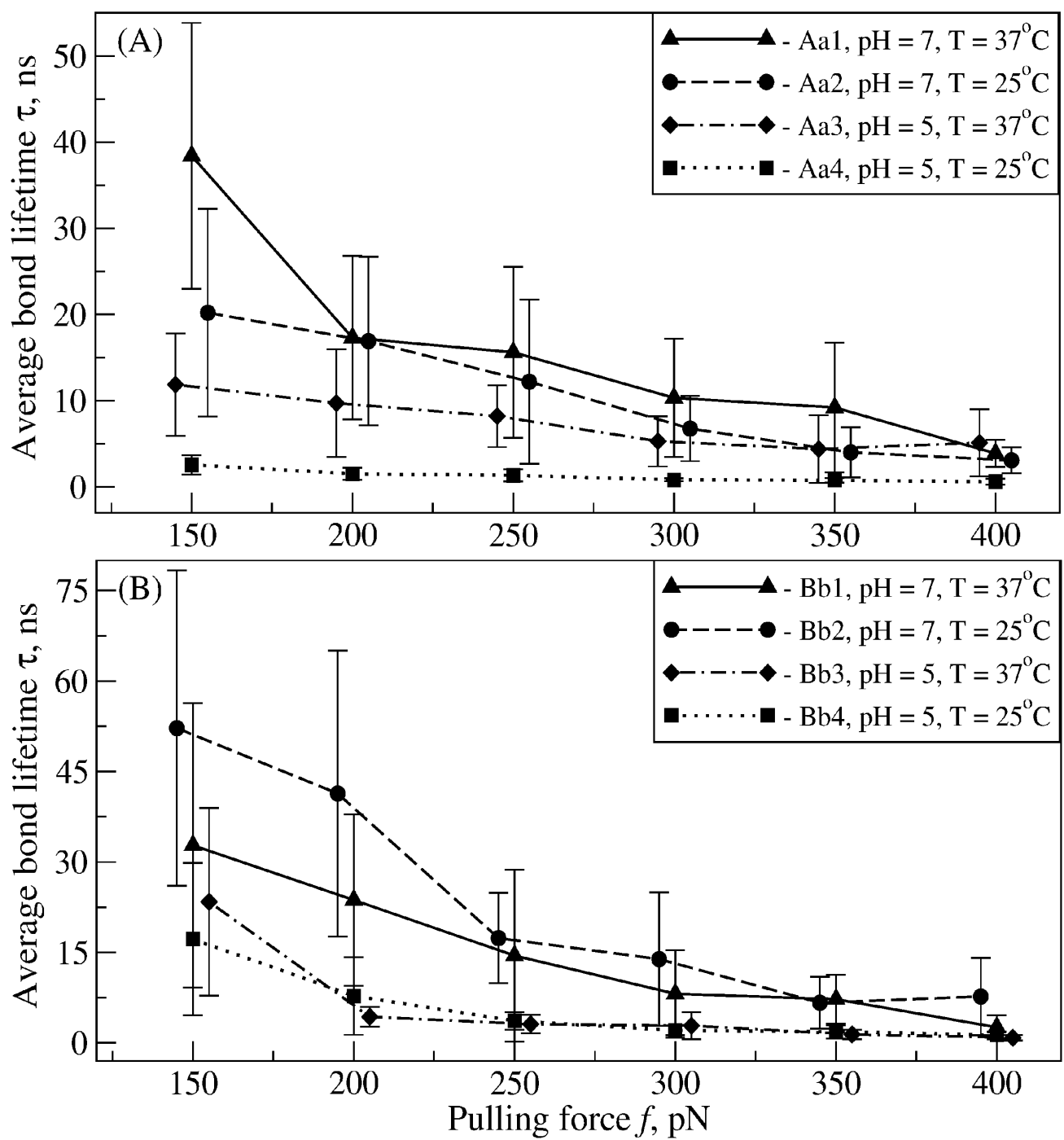

**Figure 2 (Kononova, Litvinov, Zhmurov, Alekseenko, Cheng, Agarwal, Marx, Weisel, Barsegov)**

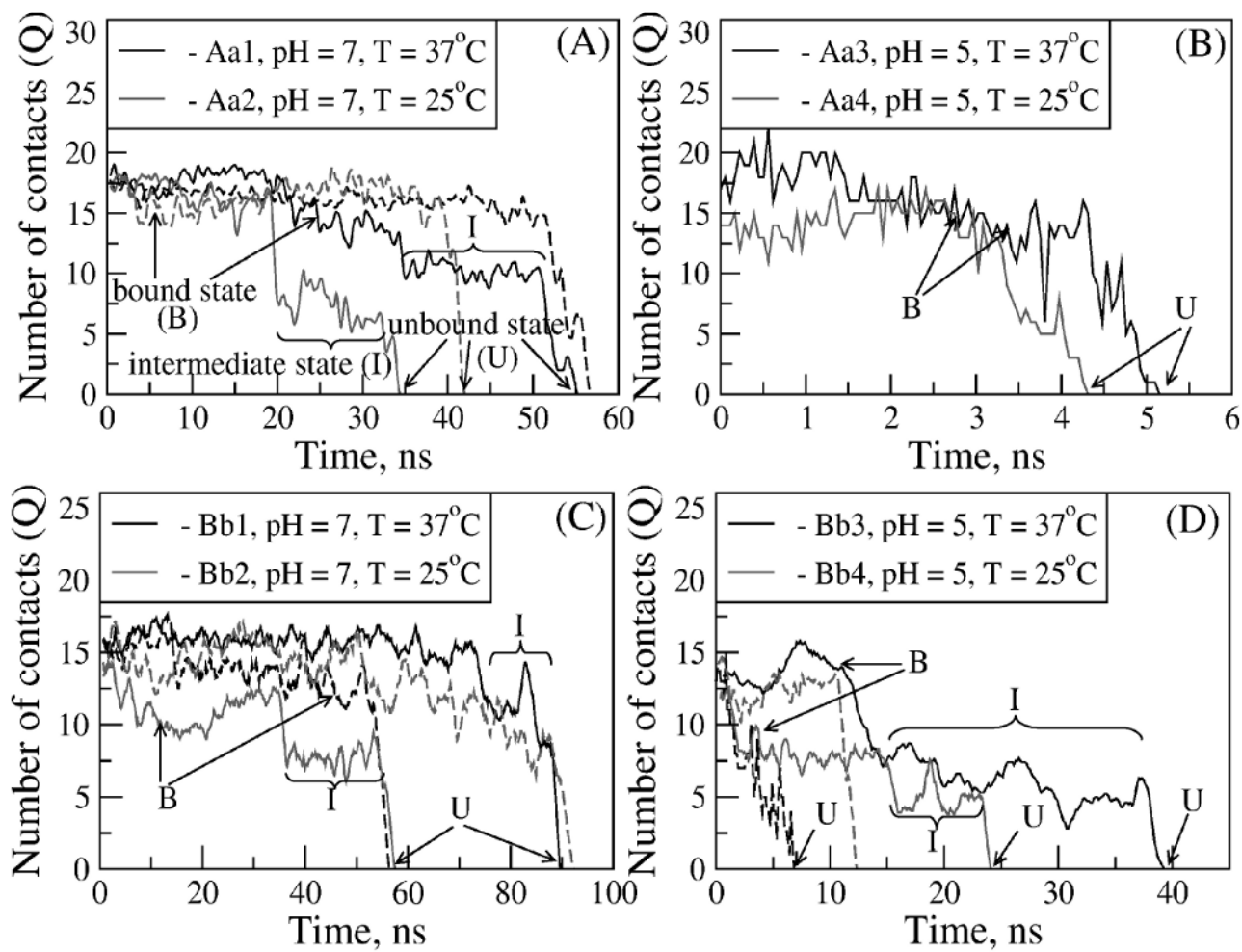

**Figure 3 (Kononova, Litvinov, Zhmurov, Alekseenko, Cheng, Agarwal, Marx, Weisel, Barsegov)**

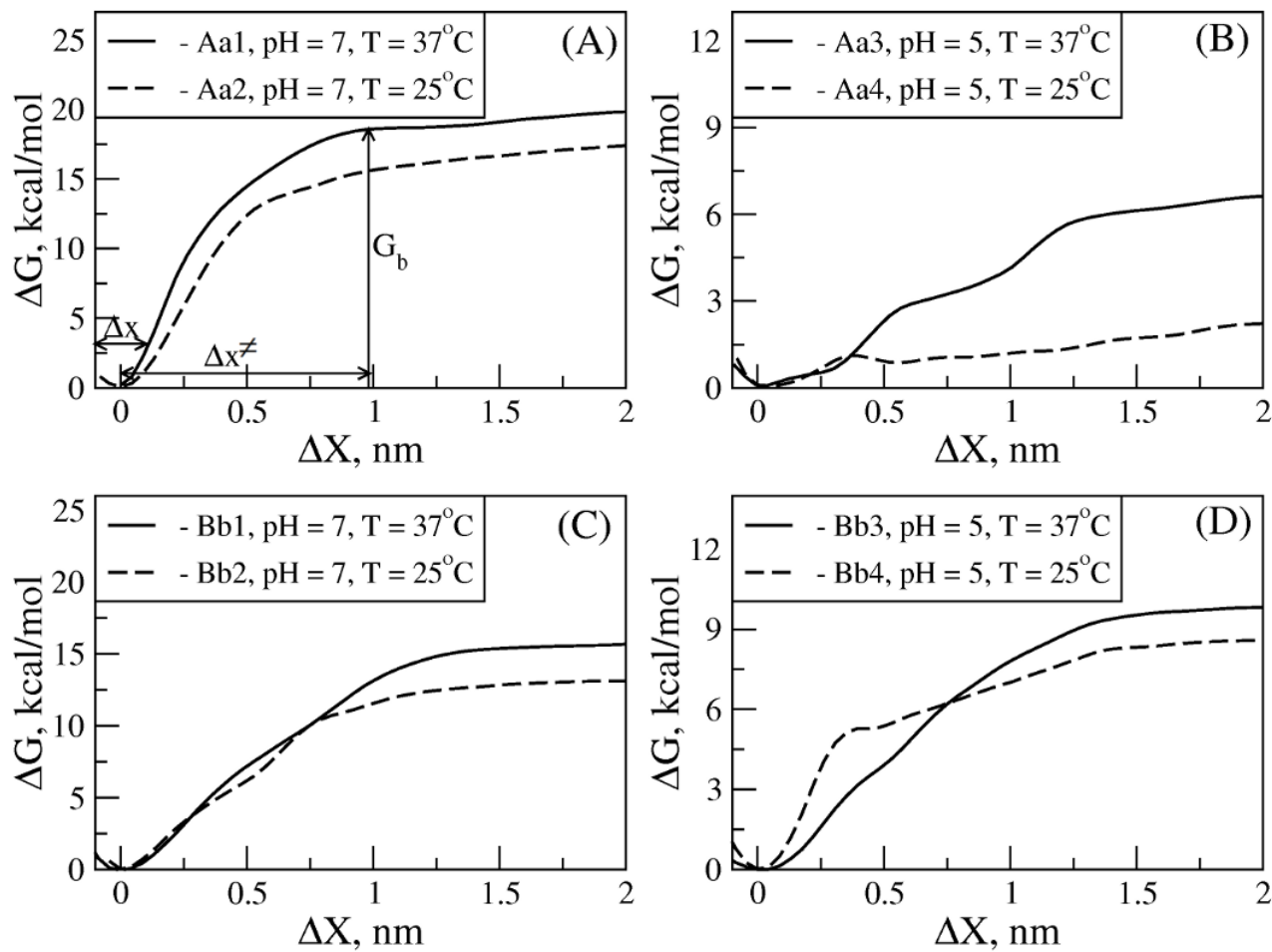

**Figure 4 (Kononova, Litvinov, Zhmurov, Alekseenko, Cheng, Agarwal, Marx, Weisel, Barsegov)**

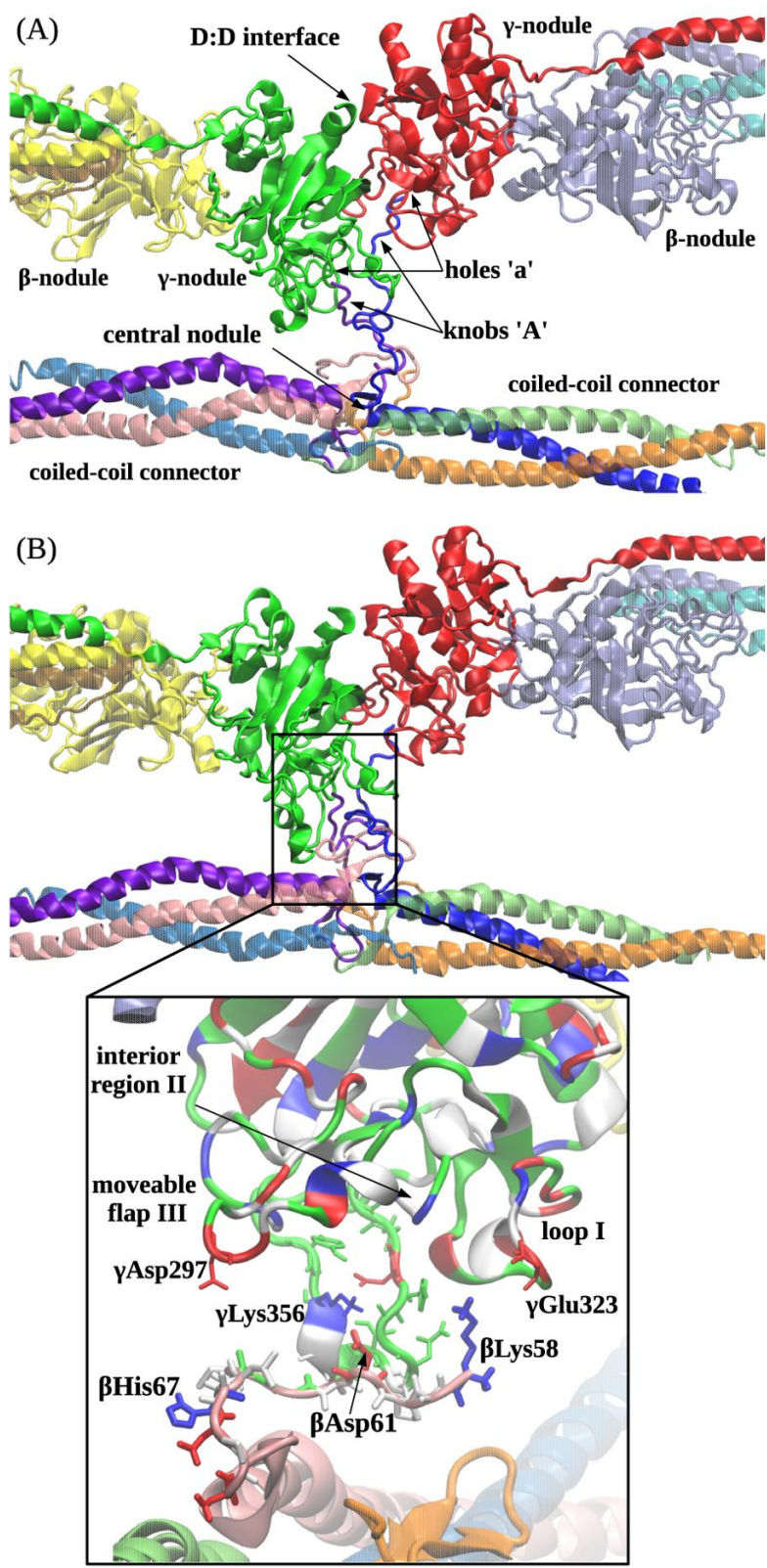

**Figure 5 (Kononova, Litvinov, Zhmurov, Alekseenko, Cheng, Agarwal, Marx, Weisel, Barsegov)**

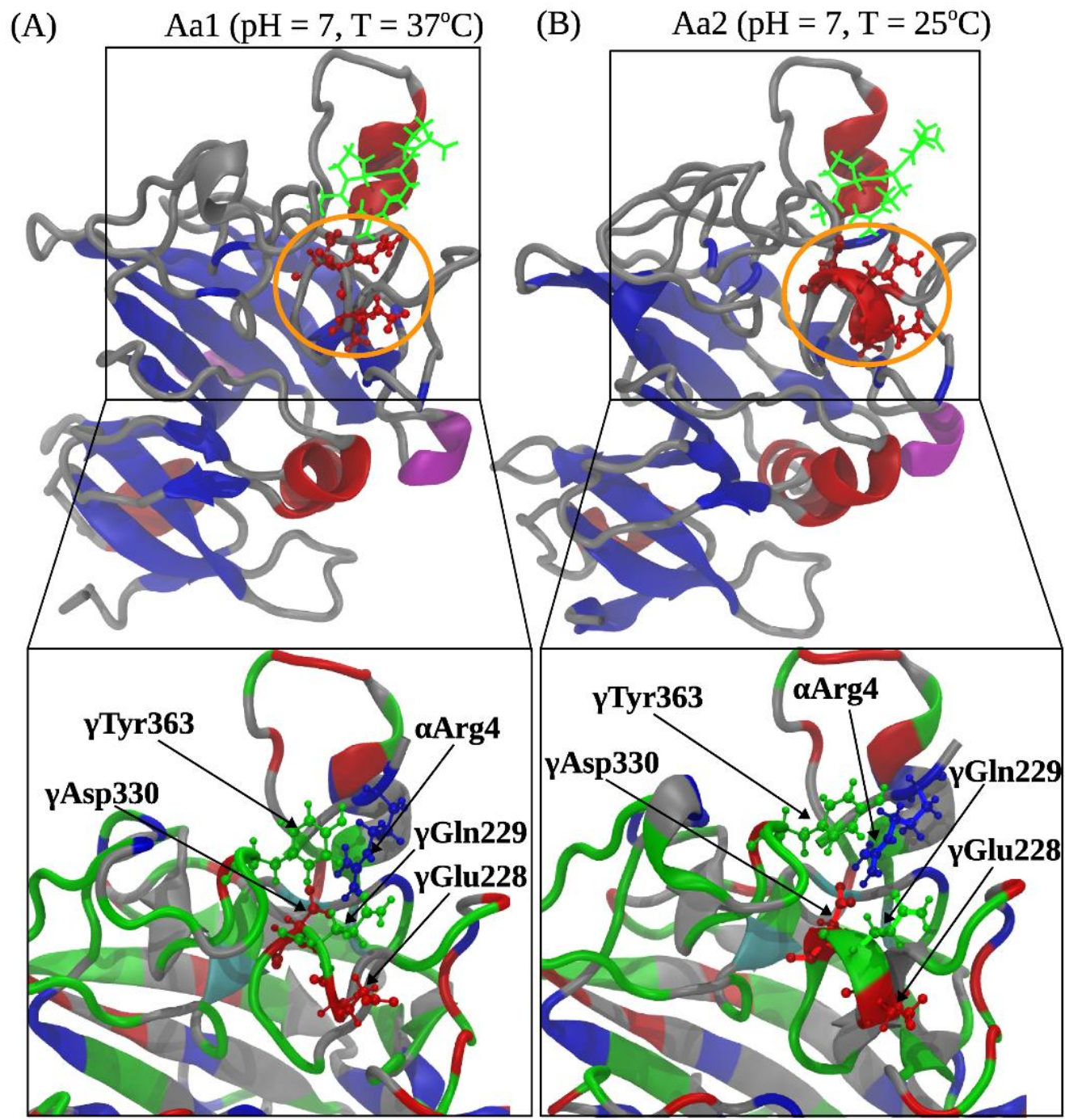

**Figure 6 (Kononova, Litvinov, Zhmurov, Alekseenko, Cheng, Agarwal, Marx, Weisel, Barsegov)**

Molecular mechanisms, thermodynamics, and kinetics of knob-hole interactions in fibrin

**Olga Kononova[1,2], Rustem I. Litvinov[3], Artem Zhmurov[1,2], Andrey Alekseenko[2], Chia Ho Cheng[1], Silvi Agarwal[1], Kenneth A. Marx[1], John W. Weisel[3], and Valeri Barsegov[1,2]**

[1]Department of Chemistry University of Massachusetts, Lowell, MA 01854

[2]Moscow Institute of Physics and Technology, Moscow Region, Russia 141700

[3]Department of Cell and Developmental Biology
University of Pennsylvania School of Medicine, Philadelphia, PA 19104

Supplemental Data

To whom correspondence should be addressed: Valeri Barsegov, Department of Chemistry, University of Massachusetts, Lowell, MA, USA, Tel: (978) 934-3661; E-mail: <u>Valeri_Barsegov@uml.edu</u>; John W. Weisel, Department of Cell and Developmental Biology, University of Pennsylvania School of Medicine, Philadelphia, PA, USA, Tel: (215) 898-3573; E-mail: <u>weisel@mail.med.upenn.edu</u>

**Essential Dynamics:** In Essential Dynamics implemented in the GROMACS package (1), the most important displacements in a biomolecular system are described by a few "essential" degrees of freedom forming the essential subspace. The remaining degrees of freedom represent less important fluctuations (2). Dynamic correlations between the positions of amino acids at time $t$, $x(t)=\{x_1(t), x_2(t),..., x_N(t)\}$ ($N$ – total number of residues), and the corresponding positions in the reference structure $x_0=\{x_1(0), x_2(0),..., x_N(0)\}$ can be expressed through the covariance matrix $C(t) = <M^{1/2}(x(t) - x_0)(M^{1/2}(x(t) - x_0))^T>$, where $<...>$ denotes the ensemble averaging and the superscript $T$ represents the transposed matrix. By construction, $C$ is a symmetric matrix, which can be diagonalized by an orthogonal transformation $T$, $C = T^TLT$, where $L$ is the matrix of eigenvalues and $T$ is the matrix of eigenvectors. For a system of $N$ particles in the three-dimensional space, there are $3N-6$ eigenvectors with non-zero eigenvalues (excluding 6 degrees of freedom for translations and rotations). The eigenvalue $L_I$ in the center-of-mass representation ($I =1,2,...,N$) is the amplitude of the $I$-th eigenvector $t_I$ along the displacement $x(t) - x_0$. The principal coordinates, $p_I(t)$, are obtained by projecting $x(t) - x_0$ onto each eigenvector: $p(t) = T^TM^{1/2}(x(t) - x_0)$. These projections are given by $y_I(t) = x_0 + M^{1/2}T^Tp_I(t)$ (2,3). To obtain the displacements of each residue along a particular eigenvector, we calculated the root-mean-square displacements $RMSD^k_I = (C_{II}/M_{II})^{1/2} = (1/M_{II}\phi_I(T^T_{Ik}L_{kl}T_{lI}))^{1/2} = (L_{kl}T_{kl}^2/M_{II})^{1/2}$, where index $k$ runs over the eigenmodes.

**Umbrella Sampling:** We calculated the potential of mean force for the force-driven dissociation of the A:a and B:b knob-hole bond, $G(X)$, as a function of the distance between the relevant residue in hole 'a' (hole 'b') and knob 'A' (knob 'B'), $X$. The potential energy of the system depends on $X$ (reaction coordinate), which gradually changes during the simulation run, and one accumulates $\left\langle \partial E \,/\, \partial X \right\rangle$ at several values of $X$. Then, $G(X)$ can be estimated as $G(X) = \int \left\langle \partial EG \,/\, \partial X \right\rangle dX$ (4,5). Because $X$ is also the distance traveled by the cantilever tip, $\partial G \,/\, \partial X$ corresponds to the restoring force $F = \kappa dX$, i.e. the force exerted on the tip by the ligand-receptor pair ($\kappa$ is the cantilever spring constant and $dX$ is the tip displacement). Hence, measurements of $F$ can be used to estimate $\partial G \,/\, \partial X$ and to resolve $G(X)$. In a single run for each model system, we performed a total of 60 steps of simulation and collected 3000 data points. In each step, the cantilever (base) was first moved by $\nabla X = 0.5$ Å and the system was equilibrated for 100 ps. Next, the values of the current tip position were sampled for 50 ps to calculate the biomolecular response $F = \partial G / \partial x$ . We used a cantilever with the spring constant $\kappa = 14$ N/m. A total of 5 independent runs have been used to perform ensemble averaging.

**Supporting References:**

**Table S1:** Model systems of the A:a and B:b knob-hole complexes for all-atom MD simulations performed under various ambient conditions (pH and temperature). Summarized for each system Aa1-Aa4 and Bb1-Bb4 are the number of independent simulation runs, and the number of neutral and positive His residues, net positive/negative charge, and number and types of counter-ions.

| System | Conditions | No. of runs | No. of His residues | Net "+/-" charge | No. of counterions |
|--------|-----------|-------------|---------------------|------------------|--------------------|
| Aa1 | pH = 7, T = 37 °C | 65 | 6 (neutral) | 29/32 | 3 Na$^+$ |
| Aa2 | pH = 7, T = 25 °C | 54 | 6 (neutral) | 29/32 | 3 Na$^+$ |
| Aa3 | pH = 5, T = 37 °C | 66 | 6 (positive) | 35/32 | 3 Cl$^-$ |
| Aa4 | pH = 5, T = 25 °C | 65 | 6 (positive) | 35/32 | 3 Cl$^-$ |
| Bb1 | pH = 7, T = 37 °C | 57 | 5 (neutral) | 34/33 | 1 Cl$^-$ |
| Bb2 | pH = 7, T = 25 °C | 60 | 5 (neutral) | 34/33 | 1 Cl$^-$ |
| Bb3 | pH = 5, T = 37 °C | 58 | 5 (positive) | 39/33 | 6 Cl$^-$ |
| Bb4 | pH = 5, T = 25 °C | 60 | 5 (positive) | 39/33 | 6 Cl$^-$ |

**Table S2.** Model systems Aa1 - Aa4 for the A:a knob-hole complex and Bb1 - Bb4 for the B:b knob-hole complex in terms of ambient conditions (pH, temperature, and salt concentration), system size (number of amino acid residues in proteins and number of water molecules), and dimensions of the solvation box.

| System | pH | T, °C | [NaCl], mM | No. of residues | No. of waters | Solvation box (x ,y, z) |
|--------|-----|-------|------------|-----------------|---------------|--------------------------|
| Aa1 | 7 | 25 | 150 | 250 | 12,066 | 48.8 Å, 50.3 Å, 67.9 Å |
| Aa2 | 7 | 37 | 150 | 250 | 12,066 | 48.8 Å, 50.3 Å, 67.9 Å |
| Aa3 | 5 | 25 | 150 | 250 | 12,066 | 48.8 Å, 50.3 Å, 67.9 Å |
| Aa4 | 5 | 37 | 150 | 250 | 12,066 | 48.8 Å, 50.3 Å, 67.9 Å |
| Bb1 | 7 | 25 | 150 | 262 | 12,338 | 45.4 Å, 80.1 Å, 72.5 Å |
| Bb2 | 7 | 37 | 150 | 262 | 12,338 | 45.4 Å, 80.1 Å, 72.5 Å |
| Bb3 | 5 | 25 | 150 | 262 | 12,336 | 45.4 Å, 80.1 Å, 72.5 Å |
| Bb4 | 5 | 37 | 150 | 262 | 12,336 | 45.4 Å, 80.1 Å, 72.5 Å |

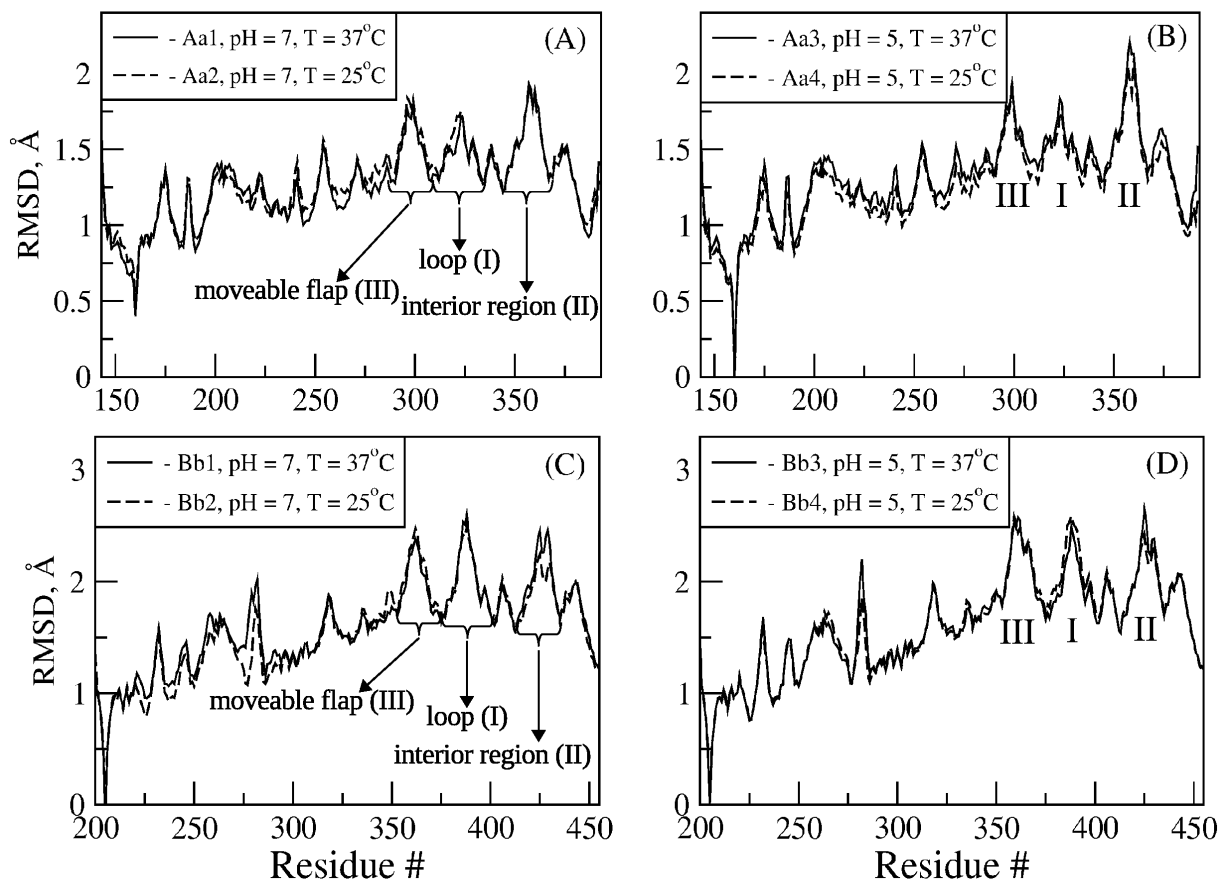

**Figure S1.** Equilibrium fluctuations of amino acid residues forming the binding interface in the holes 'a' and 'b'. Shown are the profiles of the equilibrium root-mean-square-deviations (RMSD values) for residues in hole 'a' (panels **A** and **B**) and in hole 'b' (panels **C** and **D**) co-complexed with peptides GPRP and GHRP, respectively, under different ambient conditions (pH and temperature; see Tables S1 and S2). The RMSD profiles are obtained from equilibrium simulations of the A:a knob-hole complexes - model systems Aa1 and Aa2 (panel **A**), and Aa3 and Aa4 (panel **B**), and the B:b knob-hole complexes – model systems Bb1 and Bb2 (panel **C**), and Bb3 and Bb4 (panel **D**). Amino acid residues in the binding regions I, II, and III in the holes 'a' and 'b', which participate in binding interactions with the knobs 'A' and 'B', are shown in panels **A** and **C**, respectively (see also Figure 1 in the main text).

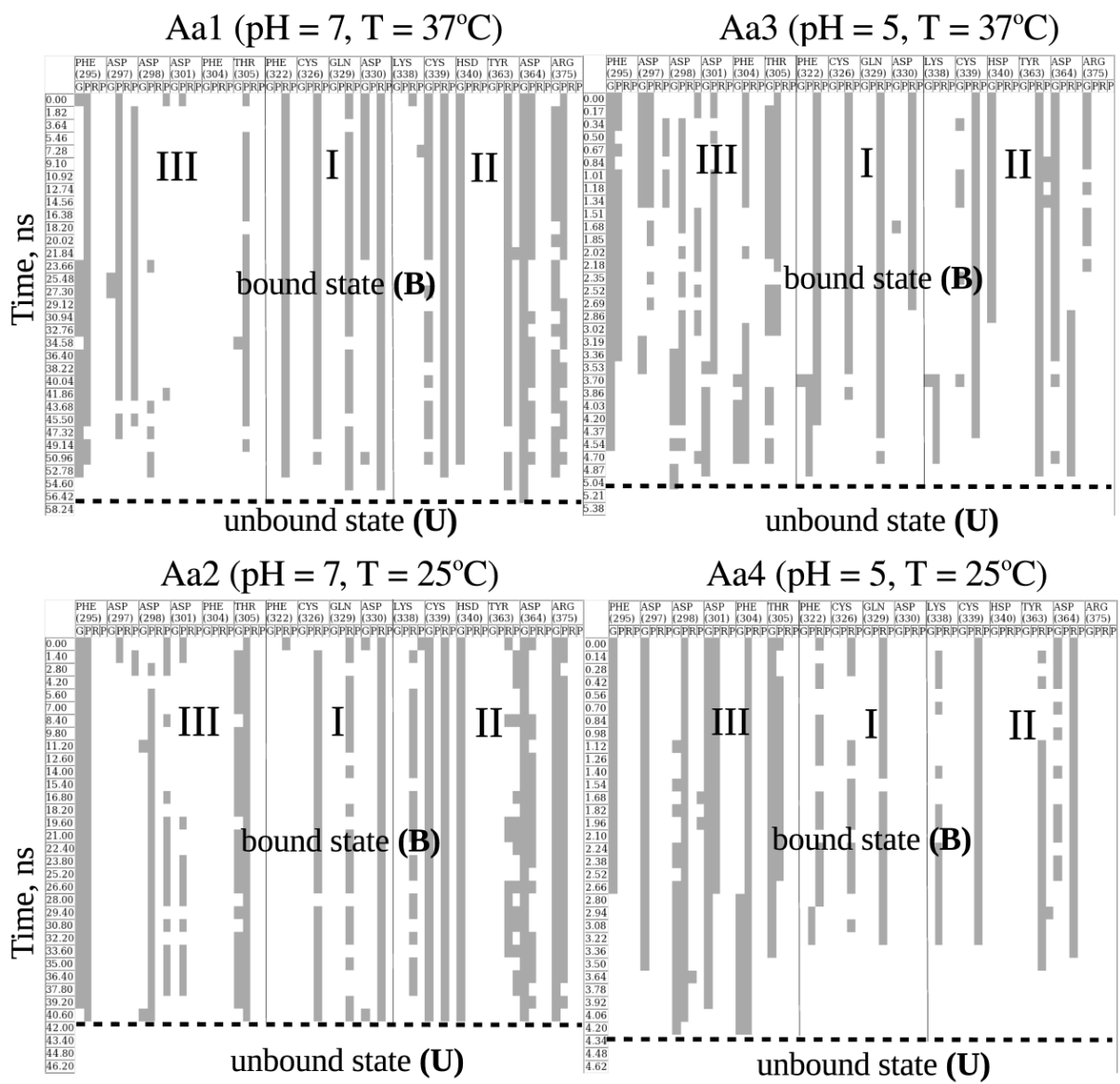

**Figure S2:** Time-dependent map of binary interactions stabilizing the A:a knob-hole complex under various ambient conditions (pH and temperature; see Tables S1 and S2). Shown are the contact maps for binary contacts between the amino acid residues in the hole 'a' and residues in the GPRP peptide (knob 'A' mimetic), obtained from the most representative trajectories of forced dissociation for model systems Aa1-Aa4. Amino acid residues in the binding regions I, II, and III in the hole 'a', which participate in binding interactions with the GPRP peptide under different environmental conditions are indicated (see Figure 1 in the main text).

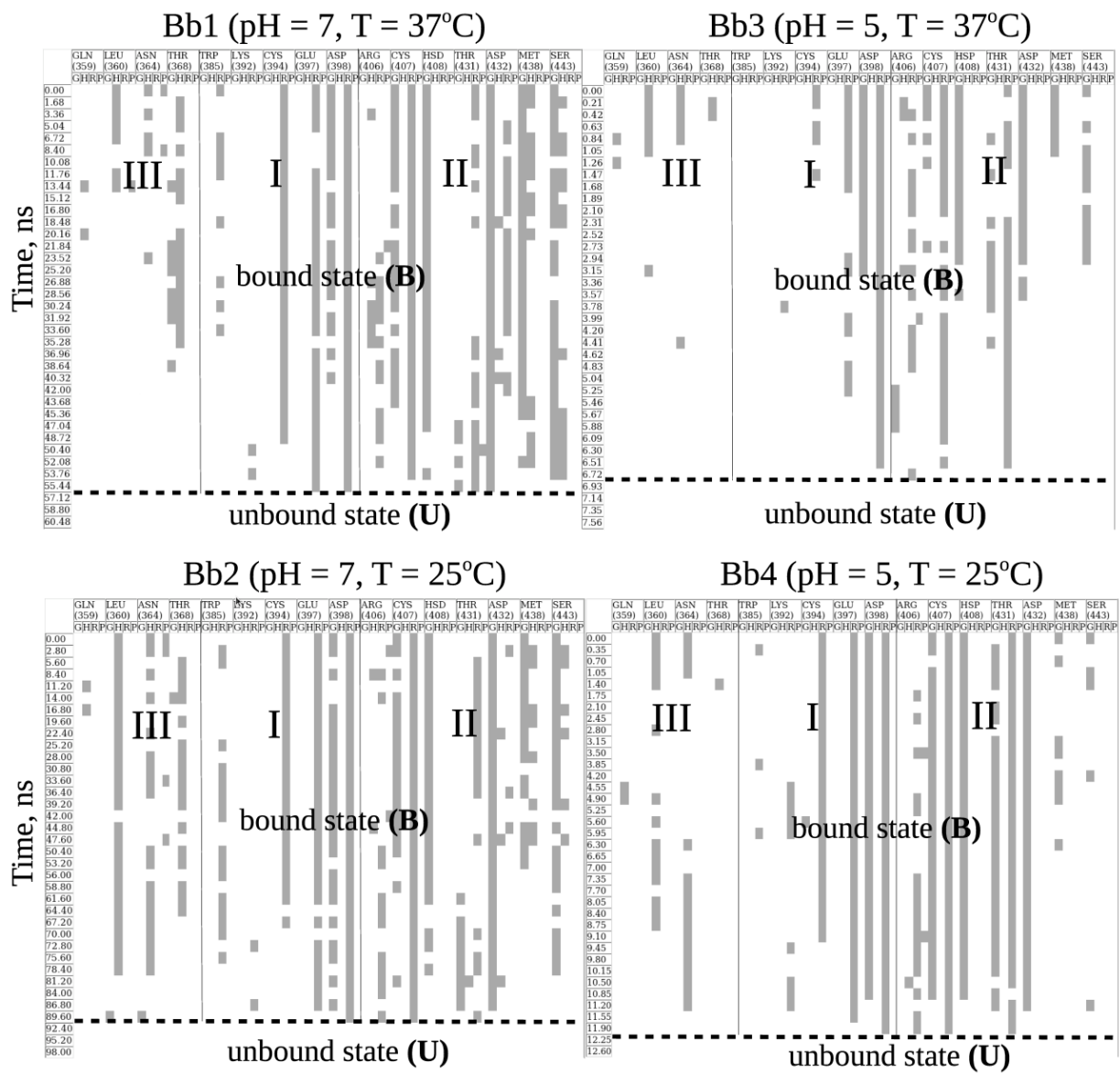

**Figure S3:** Time-dependent map of binary interactions stabilizing the B:b knob-hole complex under various ambient conditions (pH and temperature; see Tables S1 and S2). Shown are the contact maps for binary contacts between the amino acid residues in the hole 'b' and residues in the GHRP peptide (knob 'B' mimetic), obtained from the most representative trajectories of forced dissociation for model systems Bb1-Bb4. Amino acid residues in the binding regions I, II, and III in the hole 'b', which participate in binding interactions with the GHRP peptide under different solution conditions are indicated (see Figure 1 in the main text).

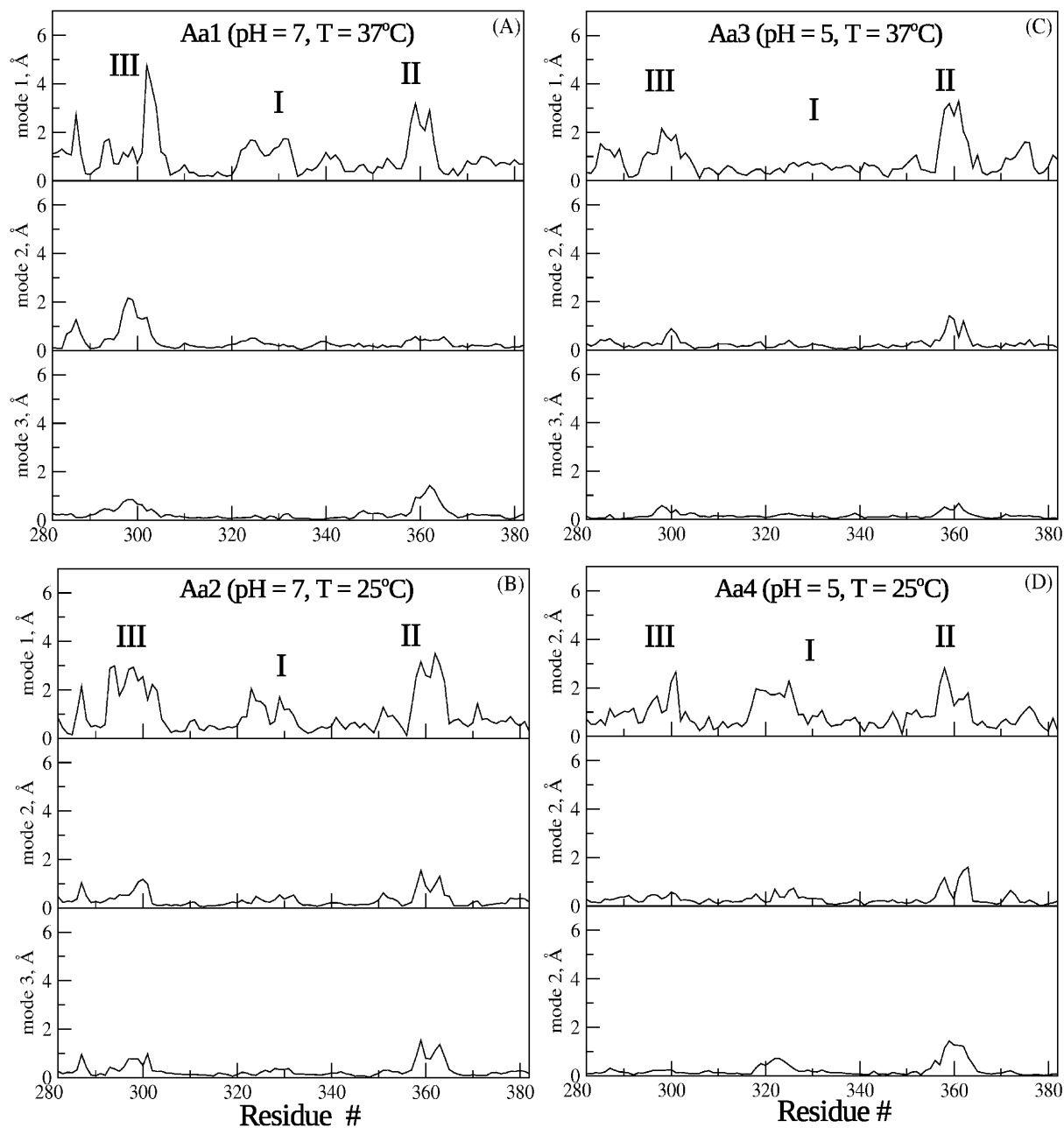

**Figure S4:** Essential dynamics underlying the forced dissociation of the A:a knob-hole complexes under various ambient conditions (pH and temperature; see Tables S1 and S2). Panels **A-D** show the profiles of the first three most important modes of displacements of binding residues in hole 'a' (mode 1, 2, and 3), which capture >90% of all the displacements accompanying forced unbinding for the A:a knob-hole complex, as a function of the residue number (γ282-γ382) for model systems Aa1-Aa4, respectively. The "essential dynamics spectra" are obtained from the most representative trajectories of forced dissociation. Corresponding to displacement peaks are amino acid residues in the binding regions I, II, and III in the hole 'a', which participate in binding interactions with the GPRP peptide (see Figure 1 in the main text).

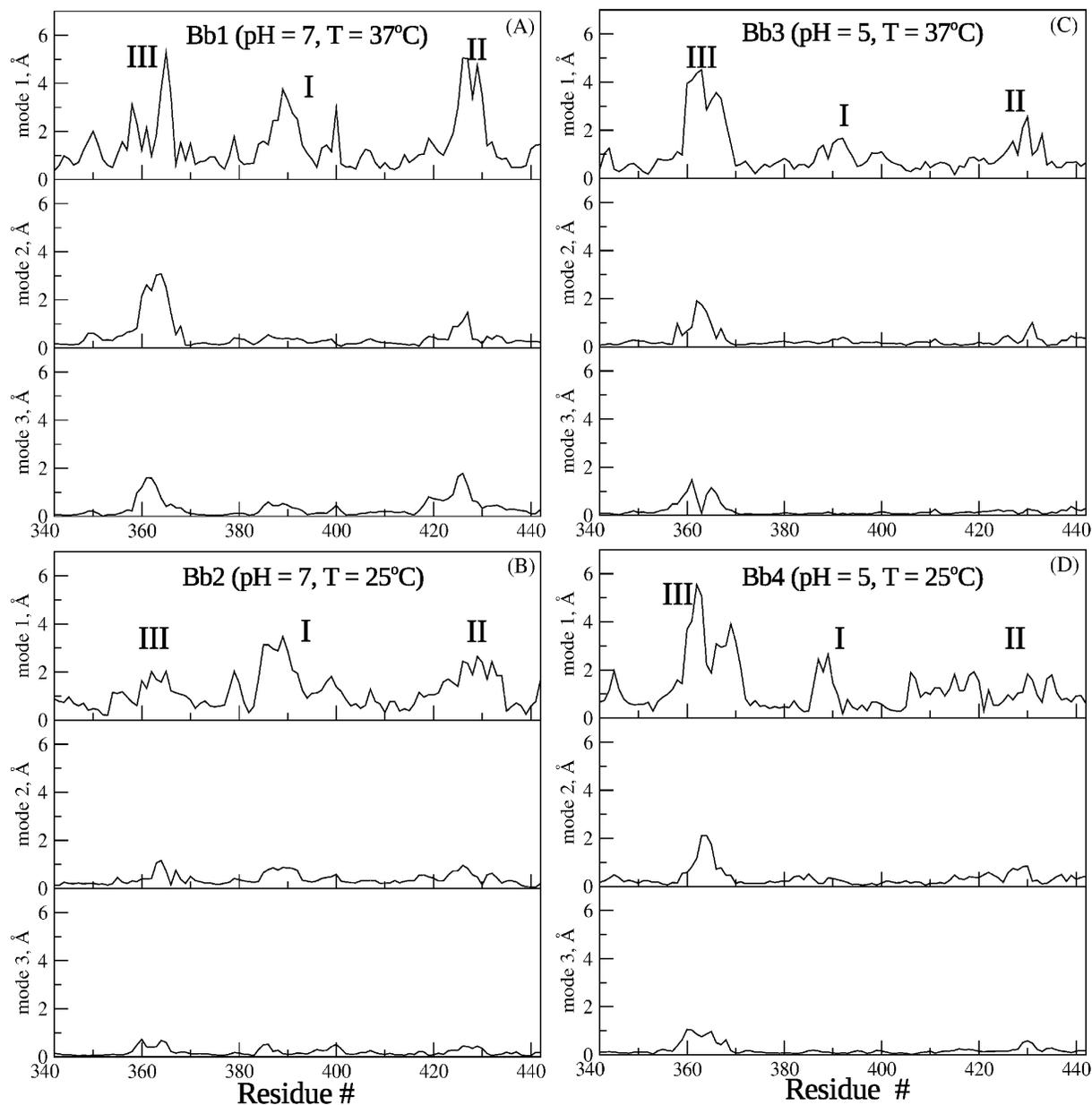

**Figure S5:** Essential dynamics underlying the forced dissociation of the B:b knob-hole complexes under various ambient conditions (pH and temperature; see Tables S1 and S2). Panels **A-D** show the profiles of the first three most important modes of displacements of binding residues in hole 'b' (mode 1, 2, and 3), which capture >90% of all the displacements accompanying forced unbinding for the B:b knob-hole complex, as a function of the residue number ($\beta$342-$\beta$442) for model systems Bb1-Bb4, respectively. The "essential dynamics spectra" are obtained from the most representative trajectories of forced dissociation. Corresponding to displacement peaks are amino acid residues in the binding regions I, II, and III in the hole 'b', which participate in binding interactions with the GHRP peptide (see Figure 1 in the main text).

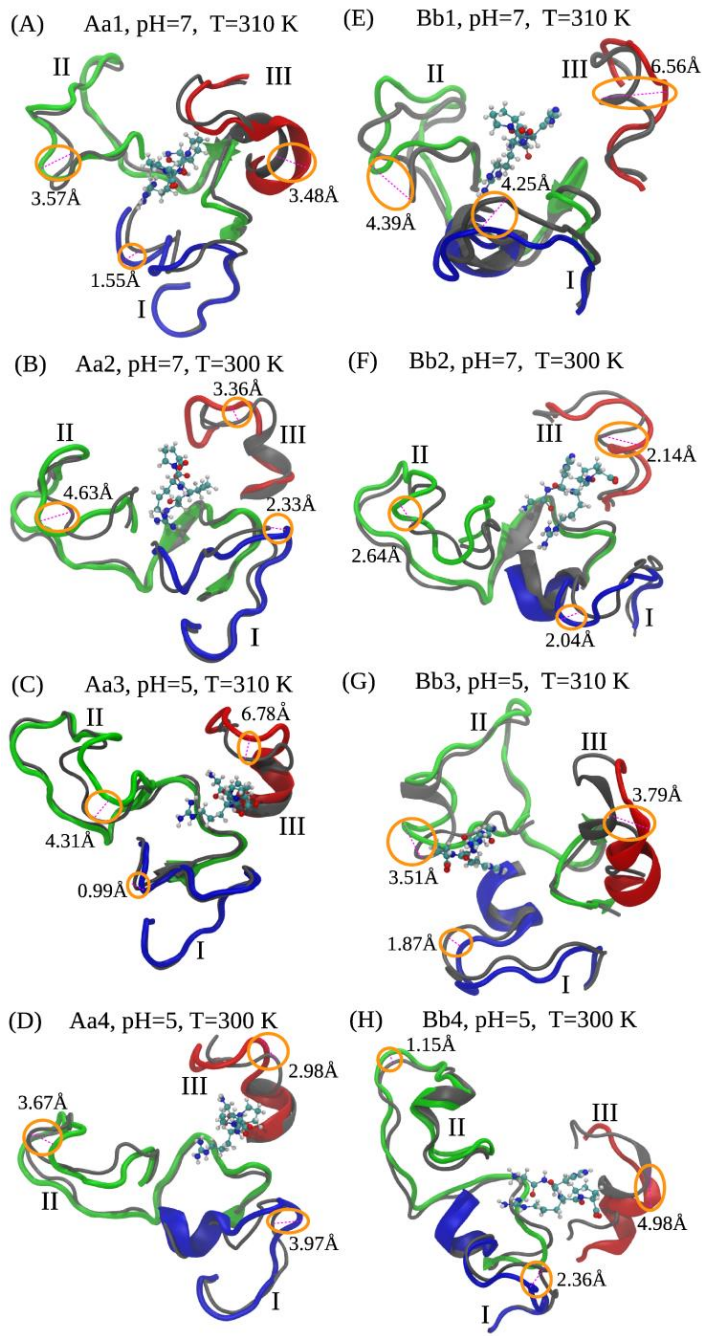

**Figure S6.** Structural transitions underlying forced dissociation of the A:a and B:b knob-hole complexes under different ambient conditions (pH and temperature; see Tables S1 and S2). Structural snapshots magnify the largest displacements in the binding interface, i.e. in binding regions I, II, and III (see Figure 1 in the main text), for model systems Aa1–Aa4 (panels **A-D**) and Bb1–Bb4 (panels **E-H**). These displacements accompany dynamic transitions from the bound state (*B*) to the globally unbound state (*U*). The types (modes) and magnitude of displacements were resolved using the Essential Dynamics approach. The three most important types of motion for the A:a and B:b knob-hole bond complexes (modes 1, 2, and 3) are profiled for different values of pH and temperature in Figures S4 and S5, respectively.